# Block-length dependent thresholds in block-sparse compressed sensing

MIHAILO STOJNIC

School of Industrial Engineering
Purdue University, West Lafayette, IN 47907
e-mail: mstojnic@purdue.edu

July 2009

### Abstract

One of the most basic problems in compressed sensing is solving an under-determined system of linear equations. Although this problem seems rather hard certain $\ell_1$-optimization algorithm appears to be very successful in solving it. The recent work of [14, 28] rigorously proved (in a large dimensional and statistical context) that if the number of equations (measurements in the compressed sensing terminology) in the system is proportional to the length of the unknown vector then there is a sparsity (number of non-zero elements of the unknown vector) also proportional to the length of the unknown vector such that $\ell_1$-optimization algorithm succeeds in solving the system. In more recent papers [78, 81] we considered the setup of the so-called **block**-sparse unknown vectors. In a large dimensional and statistical context, we determined sharp lower bounds on the values of allowable sparsity for any given number (proportional to the length of the unknown vector) of equations such that an $\ell_2/\ell_1$-optimization algorithm succeeds in solving the system. The results established in [78, 81] assumed a fairly large block-length of the block-sparse vectors. In this paper we consider the block-length to be a parameter of the system. Consequently, we then establish sharp lower bounds on the values of the allowable block-sparsity as functions of the block-length.

## 1 Introduction

In last several years the area of compressed sensing has been the subject of extensive research. Finding the sparsest solution of an under-determined system of linear equations turns out to be one of the focal points of the entire area. Recent phenomenal results of [14] and [28] rigorously proved for the first time that in certain scenarios one can solve an under-determined system of linear equations by solving a linear program in polynomial time. These breakthrough results then as expected generated enormous amount of research with possible applications ranging from high-dimensional geometry, image reconstruction, single-pixel camera design, decoding of linear codes, channel estimation in wireless communications, to machine learning, data-streaming algorithms, DNA micro-arrays, magneto-encephalography etc. (more on the compressed sensing problems, their importance, and wide spectrum of different applications can be found in excellent references [4, 12, 15, 24, 37, 58, 60, 66, 68, 70, 71, 91, 93]).

The interest of the present paper are the mathematical aspects of certain compressed sensing problems. More precisely, we will be interested in finding the sparsest solution of an under-determined system of linear equations which, as mentioned above, is one of the most fundamental problems in the compressed sensing. While the setup of this problem is fairly easy its solution is rather hard. Namely, the setup of the problem is as simple as the following: we would like to find $\mathbf{x}$ such that

$$A\mathbf{x} = \mathbf{y} \qquad (1)$$

where $A$ is an $M \times N$ ($M < N$) measurement matrix and $\mathbf{y}$ is an $M \times 1$ measurement vector. In usual compressed sensing context $\mathbf{x}$ is an $N \times 1$ unknown $K$-sparse vector (see Figure 1). This assumes that $\mathbf{x}$ has at most $K$ nonzero components (we assume ideally sparse signals; more on the so-called approximately sparse signals can be found in e.g. [21, 79, 84, 95]). In the rest of the paper we will also assume the so-called *linear* regime, i.e. we will assume that $K = \beta N$ and that the number of the measurements is $M = \alpha N$ where $\alpha$ and $\beta$ are absolute constants independent of $N$ (more on the non-linear regime, i.e. on the regime when $M$ is larger than linearly proportional to $K$ can be found in e.g. [22, 45, 46]). Since the problem given in (1) has been known for a long time there is an extensive literature related to possible ways for solving it. If one has freedom to design the measurement matrix $A$ then, clearly, a particular recovery algorithm for that design can be developed as well. As shown in [3, 59, 65], the techniques from coding theory (based on the coding/decoding of Reed-Solomon codes) can be employed to determine *any* $K$-sparse $\mathbf{x}$ in (1) for any $\alpha$ and any $\beta \leq \frac{\alpha}{2}$ in polynomial time. It is easy to see that $\beta$ can not be greater than $\frac{\alpha}{2}$ for $\mathbf{x}$ to be uniquely recoverable. Therefore in terms of recoverable sparsity in polynomial time results from [3, 59, 65] are optimal. The complexity of algorithms from [3, 59, 65] is roughly $O(N^3)$. If $A$ is designed based on the techniques related to the coding/decoding of Expander codes then the complexity of recovering $\mathbf{x}$ in (1) is $O(N)$ (see e.g. [52, 53, 94] and references therein). However, these algorithms do not allow for $\beta$ to be as large as $\frac{\alpha}{2}$.

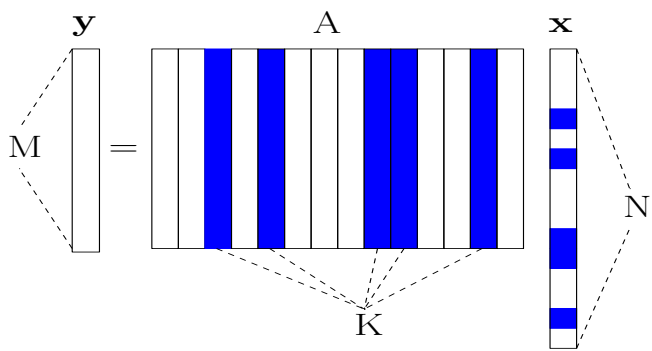

Figure 1: Model of a linear system; vector $\mathbf{x}$ is $K$-sparse

On the other hand, if there is no freedom in the choice of the matrix $A$ the problem becomes NP-hard. Two algorithms that traditionally perform well and have been the subject of an extensive research in recent years are 1) *Orthogonal matching pursuit - OMP* and 2) *Basis matching pursuit - $\ell_1$-optimization*. Both of the algorithms have advantages and disadvantages when applied to different problem scenarios. As expected a very extensive literature has been developed (especially in last several years) that covers various modifications of both algorithms so to emphasize their strengths and neutralize their flaws. However, a short assessment of their differences would be that OMP is faster while BMP can recover higher sparsity and is more resistant to system imperfections. Under certain probabilistic assumptions on the elements of the matrix $A$ it can be shown (see e.g. [62, 63, 86, 88]) that if $\alpha = O(\beta \log(\frac{1}{\beta}))$ OMP (or a slightly modified OMP) can recover $\mathbf{x}$ in (1) with complexity of recovery $O(N^2)$. On the other hand a stage-wise OMP from [36] recovers $\mathbf{x}$ in (1) with complexity of recovery $O(N \log N)$.

Since the results of this paper will in some sense be related to $\ell_1$-optimization (considered in [14, 15, 28, 34]), below we briefly recall on its definition. Basic $\ell_1$-optimization algorithm (more on adaptive versions of basic $\ell_1$-optimization can be found in e.g. [16, 19, 76]) finds $\mathbf{x}$ in (1) by solving the following problem

$$\begin{aligned} \min \quad & \|\mathbf{x}\|_1 \\ \text{subject to} \quad & A\mathbf{x} = \mathbf{y}. \end{aligned} \qquad (2)$$

(Instead of $\ell_1$-optimization one can employ $\ell_q$-optimization, $0 < q < 1$, which essentially means that instead of norm 1 one can use norm $q$ in (1). However the resulting problem becomes non-convex. A good overview of that approach can be found in e.g. [26, 43, 48–50, 75] and references therein.) Quite remarkably, in [15] the authors were able to show that if $\alpha$ and $N$ are given, the matrix $A$ is given and satisfies a special property called the restricted isometry property (RIP), then any unknown vector $\mathbf{x}$ with no more than $K = \beta N$ (where $\beta$ is an absolute constant dependent on $\alpha$ and explicitly calculated in [15]) non-zero elements can be recovered by solving (2). As expected, this assumes that $\mathbf{y}$ was in fact generated by that $\mathbf{x}$ and given to us. The case when the available measurements are noisy versions of $\mathbf{y}$ is also of interest [14, 15, 51, 92]. We mention in passing that the recent popularity of $\ell_1$-optimization in compressed sensing is significantly due to its robustness with respect to noisy measurements. (Of course, the main reason for its popularity is its ability to solve (1) for a very wide range of matrices $A$; more on this remarkable universality phenomenon the interested reader can find in [33].)

Since the RIP condition played a crucial role in proving technique of [14, 15] having the matrix $A$ satisfy the RIP condition is fundamentally important. (More on the importance of the RIP condition can be found in [13]). Designing deterministic matrices for which the RIP condition would hold as well as checking if it holds for any given matrix is a very hard problem. However, for several classes of random matrices (e.g., matrices with i.i.d. zero mean Gaussian, Bernoulli or even general Sub-gaussian components) it turns out that for certain dimensions of the system the RIP condition is satisfied with overwhelming probability [1, 5, 15, 73]. On the other hand, it should also be pointed out that the RIP is only a *sufficient* condition for $\ell_1$-optimization to produce the solution of (1). In turn this means that an analysis of $\ell_1$-optimization success is not required to rely on it.

In fact, the final results and brilliant analysis of [27, 28] do not rely on the validity of the RIP condition. Namely, in [27, 28] the author considers polytope obtained by projecting the regular $N$-dimensional cross-polytope using the matrix $A$. It turns out that a *necessary and sufficient* condition for (2) to produce the solution of (1) for any given $\mathbf{x}$ is that this polytope associated with the matrix $A$ is $K$-neighborly [27–30]. Using the results of [2, 10, 72, 90], it is further shown in [28], that if the matrix $A$ is a random $m \times n$ ortho-projector matrix then with overwhelming probability polytope obtained projecting the standard $N$-dimensional cross-polytope by $A$ is $K$-neighborly. The precise relation between $M$ and $K$ in order for this to happen is characterized in [27, 28] as well.

It should be noted that one usually considers success of (2) in finding solution of (1) for *any* given $\mathbf{x}$. It is also of interest to consider success of (2) in finding solution of (1) for *almost any* given $\mathbf{x}$. To make a distinction between these two cases we will in the following section recall on several important definitions from [28, 29, 31].

Before proceeding further we first in the following section introduce the so-called block-sparse signals that will be the central topic of this paper. Immediately afterwards we also describe a polynomial algorithm for their efficient recovery.

## 2 Block-sparse signals and $\ell_2/\ell_1$-algorithm

What we described in the previous section is the standard compressed sensing setup. Such a setup does not assume any special structure on the unknown $K$-sparse signal $\mathbf{x}$. However one may encounter applications when the signal $\mathbf{x}$ in addition to being sparse has a certain structure. The so-called block-sparse signals were introduced and its applications and recovery algorithms were investigated in [4, 17, 38–40, 44, 65, 78, 81, 83]. A related problem of recovering jointly sparse signals and its applications were considered in [6, 9, 18, 23, 41, 61, 64, 85, 87, 89, 91, 97, 98] and references therein (more on different types of a priori known signal structure can also be found in [55, 56, 96]). In all these cases one attempts to improve the recoverability potential of the standard algorithms described in the previous section by incorporating the knowledge of the signal structure.

In this paper we will be interested in further investigating the so-called block-sparse compressed sensing problems [4, 40, 65, 78, 81, 83]. To introduce block-sparse signals and facilitate the subsequent exposition we will assume that integers $N$ and $d$ are chosen such that $n = \frac{N}{d}$ is an integer and it represents the total number of blocks that $\mathbf{x}$ consists of. Clearly $d$ is the length of each block. Furthermore, we will assume that $m = \frac{M}{d}$ is an integer as well and that $\mathbf{X}_i = \mathbf{x}_{(i-1)d+1:id}, 1 \leq i \leq n$ are the $n$ blocks of $\mathbf{x}$ (see Figure 2). Then we will call any signal $\mathbf{x}$ k-block-sparse if its at most $k = \frac{K}{d}$ blocks $\mathbf{X}_i$ are non-zero (non-zero

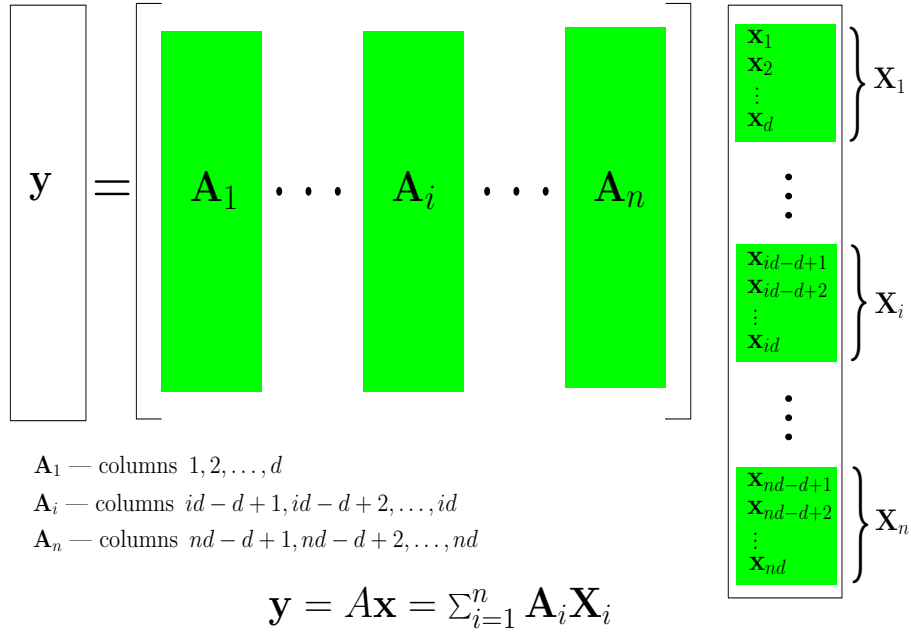

Figure 2: Block-sparse model

block is a block that is not a zero block; zero block is a block that has all elements equal to zero). Since $k$-block-sparse signals are $K$-sparse one could then use (2) to recover the solution of (1). While this is possible, it clearly uses the block structure of $\mathbf{x}$ in no way. To exploit the block structure of $\mathbf{x}$ in [83] the following polynomial algorithm (essentially a combination of $\ell_2$ and $\ell_1$ optimizations) was considered (see

also e.g. [4, 39, 89, 97, 98])

$$\begin{aligned}\min \quad & \sum_{i=1}^{n}\|\mathbf{x}_{(i-1)d+1:id}\|_2 \\ \text{subject to} \quad & A\mathbf{x} = \mathbf{y}.\end{aligned} \qquad (3)$$

Extensive simulations in [83] demonstrated that as $d$ grows the algorithm in (3) significantly outperforms the standard $\ell_1$. The following was shown in [83] as well: let $A$ be an $M \times N$ matrix with a basis of null-space comprised of i.i.d. Gaussian elements; if $\alpha = \frac{M}{N} \to 1$ then there is a constant $d$ such that all $k$-block-sparse signals $\mathbf{x}$ with sparsity $K \leq \beta N, \beta \to \frac{1}{2}$, can be recovered with overwhelming probability by solving (3). The precise relation between $d$ and how fast $\alpha \longrightarrow 1$ and $\beta \longrightarrow \frac{1}{2}$ was quantified in [83] as well. In [78, 81] we extended the results from [83] and obtained the values of the recoverable block-sparsity for any $\alpha$, i.e. for $0 \leq \alpha \leq 1$. More precisely, for any given constant $0 \leq \alpha \leq 1$ we in [78, 81] determined a constant $\beta = \frac{K}{N}$ such that for a sufficiently large $d$ (3) with overwhelming probability recovers any $k$-block-sparse signal with sparsity less then $K$. (Under overwhelming probability we in this paper assume a probability that is no more than a number exponentially decaying in $N$ away from 1.)

Clearly, for any given constant $\alpha \leq 1$ there is a maximum allowable value of the constant $\beta$ such that (3) finds solution of (1) with overwhelming probability for *any* $\mathbf{x}$. This maximum allowable value of the constant $\beta$ is called the *strong threshold* (see [27, 28]). We will denote the value of the strong threshold by $\beta_s$. Similarly, for any given constant $\alpha \leq 1$ one can define the *sectional threshold* as the maximum allowable value of the constant $\beta$ such that (3) finds the solution of (1) with overwhelming probability for *any* $\mathbf{x}$ with a given fixed location of non-zero blocks (see [27, 28]). In a similar fashion one can then denote the value of the sectional threshold by $\beta_{sec}$. Finally, for any given constant $\alpha \leq 1$ one can define the *weak threshold* as the maximum allowable value of the constant $\beta$ such that (3) finds the solution of (1) with overwhelming probability for *any* $\mathbf{x}$ with a given fixed location of non-zero blocks and given fixed directions of non-zero block vectors $\mathbf{X}_i$ (see [27, 28]). In a similar fashion one can then denote the value of the weak threshold by $\beta_w$.

While [78, 81] provided fairly sharp threshold values they had done so in a somewhat asymptotic sense. Namely, the analysis presented in [78, 81] assumed fairly large values of block-length $d$. As such the analysis in [78, 81] then provided an ultimate performance limit of $\ell_2/\ell_1$-optimization rather than its performance characterization as a function of a particular fixed block-length. In this paper we extend the results from [78, 81] so that the threshold values are now functions of a fixed block-length $d$. Our analysis will use some ingredients of the analysis presented in [78, 81]. However, significantly more precise estimates of certain quantities will be necessary to account for a fixed block-length. These estimates will be obtained in a fashion similar to the one presented in [82]. In addition to the strong thresholds (which were the main concern of [78, 81]), we will also determine attainable values for the sectional and weak thresholds as functions of a fixed block-length $d$ for the entire range of $\alpha$, i.e. for any $0 < \alpha \leq 1$.

We organize the rest of the paper in the following way. In Section 3 we introduce two key theorems that will be the heart of our subsequent analysis. In Section 4 we determine the values of the strong, sectional, and weak thresholds for a given block-length $d$ under the assumption that the null-space of the matrix $A$ is uniformly distributed in the Grassmanian. In Section 5 we determine the asymptotic values of the strong, sectional, and weak thresholds assuming large block length $d$. In Section 6 we present the results of the conducted numerical experiments and finally, in Section 7 we discuss obtained results and possible directions for future work.

## 3   Null-space and escape through a mesh theorems

In this section we introduce two useful theorems that will be of key importance in our subsequent analysis. First we recall on a null-space characterization of the matrix $A$ which establishes a guarantee that the solutions of (1) and (3) coincide. The following theorem from [78, 81, 83] provides this characterization. Set $\mathcal{K}$ to be the set of all subsets of size $k$ of $\{1, 2, \ldots, n\}$; also if $\kappa \subset \mathcal{K}$ then $\kappa^c = \{1, 2, \ldots, n\} \setminus \kappa$.

**Theorem 1.** *( [83]) Assume that $A$ is a $dm \times dn$ measurement matrix, $\mathbf{y} = A\mathbf{x}$ and $\mathbf{x}$ is $k$-block-sparse. Then the solutions of (3) and (1) coinside if and only if for all nonzero $\mathbf{w} \in R^{dn}$ where $A\mathbf{w} = 0$ and all $\kappa \in \mathcal{K}$*

$$\sum_{i \in \kappa} ||\mathbf{W}_i||_2 < \sum_{i \in \kappa^c} ||\mathbf{W}_i||_2 \qquad (4)$$

*where $\mathbf{W}_i = (\mathbf{w}_{(i-1)d+1}, \mathbf{w}_{(i-1)d+2}, \ldots, \mathbf{w}_{id})^T$, $i = 1, 2, \ldots, n$.*

The following three remarks seem to be in order.

**Remark 1:** The following simplification of the previous theorem is also well-known. Let $\mathbf{w} \in \mathbf{R}^n$ be such that $A\mathbf{w} = 0$. Further, let $\mathbf{W}_{(norm)} = (||\mathbf{W}_1||_2, ||\mathbf{W}_2||_2, \ldots, ||\mathbf{W}_n||_2)^T$ and let $|\mathbf{W}_{(norm)}|_{(i)}$ be the $i$-th smallest of the elements of $\mathbf{W}_{(norm)}$. Set $\tilde{\mathbf{W}} = (|\mathbf{W}_{(norm)}|_{(1)}, |\mathbf{W}_{(norm)}|_{(2)}, \ldots, |\mathbf{W}_{(norm)}|_{(n)})^T$. If $(\forall \mathbf{w} | A\mathbf{w} = 0) \sum_{i=n-k+1}^{n} \tilde{\mathbf{W}}_i \leq \sum_{i=1}^{n-k} \tilde{\mathbf{W}}_i$, where $\tilde{\mathbf{W}}_i$ is the $i$-th element of $\tilde{\mathbf{W}}$, then the solutions of (1) and (3) coincide.

**Remark 2:** Characterization given in the previous theorem (and proven in [83]) is a mere analogue to the similar characterizations related to the equivalence of (1) and (2) from e.g. [32, 35, 42, 57, 80, 83, 95, 99].

If instead of $\ell_1$ one, for example, uses an $\ell_q$-optimization ($0 < q < 1$) in (2) then characterizations similar to the ones from [32, 35, 42, 57, 83, 95, 99] can be derived as well [48–50]. In a similar fashion one could then derive an equivalent to the previous theorem for the $\ell_2/\ell_q$-optimization, $0 < q < 1$.

**Remark 3:** Checking if the condition given in the above theorem is satisfied for a given matrix $A$ is a very important and difficult problem. Although it is not the main topic of the present paper, we do mention in passing that a possible approximate way of solving it would be a generalization of results from e.g. [25, 54].

Clearly, if one can construct the matrix $A$ such that (4) holds then the solution of (3) would be the solution of (1). If one assumes that $m$ and $k$ are proportional to $n$ (the case of our interest in this paper) then the construction of the deterministic matrices $A$ that would satisfy (4) is not an easy task. However, if one turns to random matrices this appears to be significantly easier. In the following sections we will show that this is indeed possible for a particular type of random matrices.

More precisely, as we have already hinted earlier, we will consider the random matrices $A$ that have the null-space uniformly distributed in the Grassmanian. The following phenomenal result from [47] that relates to such matrices will be one of key ingredients in the analysis that will follow.

**Theorem 2.** *( [47] Escape through a mesh) Let $S$ be a subset of the unit Euclidean sphere $S^{dn-1}$ in $R^{dn}$. Let $Y$ be a random $d(n-m)$-dimensional subspace of $R^{dn}$, distributed uniformly in the Grassmanian with respect to the Haar measure. Let*

$$w(S) = E \sup_{\mathbf{w} \in S}(\mathbf{h}^T \mathbf{w}) \tag{5}$$

*where $\mathbf{h}$ is a random column vector in $R^{dn}$ with i.i.d. $\mathcal{N}(0,1)$ components, $\mathbf{w}$ is a $dn$-dimensional column vector from $S$, and $\mathbf{h}^T$ is the transpose of $\mathbf{h}$. Assume that $w(S) < \left(\sqrt{dm} - \frac{1}{4\sqrt{dm}}\right)$. Then*

$$P(Y \cap S = 0) > 1 - 3.5 e^{-\frac{\left(\sqrt{dm} - \frac{1}{4\sqrt{dm}} - w(S)\right)^2}{18}}. \tag{6}$$

**Remark**: Gordon's original constant $3.5$ was substituted by $2.5$ in [74]. Both constants are fine for our subsequent analysis.

## 4 Probabilistic analysis of the null-space characterizations

In this section we probabilistically analyze validity of the null-space characterization given in Theorem 1. In the first subsection of this section we will show how one can obtain the values of the strong threshold $\beta_s$ for the entire range $0 \leq \alpha \leq 1$ based on such an analysis. In the later two subsections we will extend the strong threshold analysis and obtain the values of the sectional and weak thresholds.

## 4.1 Strong threshold

As masterly noted in [74] Theorem 2 can be used to probabilistically analyze (4) (and as we will see later in the paper, many of its variants). Namely, let $S$ in (5) be

$$S_s = \{\mathbf{w} \in S^{dn-1} | \sum_{i=n-k+1}^{n} \tilde{\mathbf{W}}_i \leq \sum_{i=1}^{n-k} \tilde{\mathbf{W}}_i\} \tag{7}$$

where as earlier the notation $\tilde{\mathbf{W}}$ is used to denote the vector obtained by sorting the elements of $\mathbf{W}_{(norm)}$ in non-decreasing order (essentially, $\tilde{\mathbf{W}}$ is a vector obtained by sorting magnitudes of blocks $\mathbf{W}_i$ in non-decreasing order). Also, here and in an analogous fashion in the later sections of the paper, we assume that $k$ is such that there is an $\alpha$, $0 < \alpha \leq 1$, such that the solutions of (1) and (3) coincide. Let $Y$ be a $d(n-m)$ dimensional subspace of $R^{dn}$ uniformly distributed in Grassmanian. Furthermore, let $Y$ be the null-space of $A$. Then as long as $w(S_s) < \left(\sqrt{dm} - \frac{1}{4\sqrt{dm}}\right)$, $Y$ will miss $S_s$ (i.e. (4) will be satisfied) with probability no smaller than the one given in (6). More precisely, if $\alpha = \frac{m}{n}$ is a constant (the case of interest in this paper), $n, m$ are large, and $w(S_s)$ is smaller than but proportional to $\sqrt{dm}$ then $P(Y \cap S_s = 0) \longrightarrow 1$. This in turn is equivalent to having

$$P(\forall \mathbf{w} \in R^{dn} | A\mathbf{w} = 0, \sum_{i=n-k+1}^{n} \tilde{\mathbf{W}}_i \leq \sum_{i=1}^{n-k} \tilde{\mathbf{W}}_i) \longrightarrow 1$$

which according to Theorem 1 (or more precisely according to remark 1 after Theorem 1) means that the solutions of (1) and (3) coincide with probability 1. For any given value of $\alpha \in (0,1)$ a threshold value of $\beta$ can then be determined as a maximum $\beta$ such that $w(S_s) < \left(\sqrt{dm} - \frac{1}{4\sqrt{dm}}\right)$. That maximum $\beta$ will be exactly the value of the strong threshold $\beta_s$. If one is only concerned with finding a possible value for $\beta_s$ it is easy to note that instead of computing $w(S_s)$ it is sufficient to find its an upper bound. However, to determine as good values of $\beta_s$ as possible, the upper bound on $w(S_s)$ should be as tight as possible. The main contribution of this work will be a fairly precise estimate of $w(S_s)$.

In the following subsections we present a way to get such an estimate. To simplify the exposition we first set $w(\mathbf{h}, S_s) = \max_{\mathbf{w} \in S_s}(\mathbf{h}^T \mathbf{w})$. In order to upper-bound $w(S_s)$ we will first in Subsection 4.1.1 determine an upper bound $B_s$ on $w(\mathbf{h}, S_s)$. The expected value with respect to $\mathbf{h}$ of such an upper bound will be an upper bound on $w(S_s)$. In Subsection 4.1.2 we will compute an upper bound on that expected value, i.e. we will compute an upper bound on $E(B_s)$. That quantity will be an upper bound on $w(S_s)$ since according to the following $E(B_s)$ is an upper bound on $w(S_s)$

$$w(S_s) = Ew(\mathbf{h}, S_s) = E(\max_{\mathbf{w} \in S_s}(\mathbf{h}^T \mathbf{w})) \leq E(B_s). \tag{8}$$

### 4.1.1 Upper-bounding $w(\mathbf{h}, S_s)$

Let $\mathbf{H}_i = (\mathbf{h}_{(i-1)d+1}, \mathbf{h}_{(i-1)d+2}, \ldots, \mathbf{h}_{id})^T$, $i = 1, 2, \ldots, n$. From the definition of set $S_s$ given in (7) it easily follows that if $\mathbf{w}$ is in $S_s$ then any vector obtain from $\mathbf{w}$ by rotating (essentially multiplying by orthogonal matrices) any subset of its blocks $\mathbf{W}_i, 1 \leq i \leq n$, in any direction is also in $S_s$. The directions of vectors $\mathbf{W}_i, 1 \leq i \leq n$, can therefore be chosen so that they match the directions of vectors $\mathbf{H}_i, 1 \leq i \leq n$ of the corresponding blocks in $\mathbf{h}$. We then easily have

$$w(\mathbf{h}, S_s) = \max_{\mathbf{w} \in S_s}(\mathbf{h}^T \mathbf{w}) = \max_{\mathbf{w} \in S_s} \sum_{i=1}^{n} |\mathbf{h}_i \mathbf{w}_i| = \max_{\mathbf{w} \in S_s} \sum_{i=1}^{n} \|\mathbf{H}_i\|_2 \|\mathbf{W}_i\|_2. \tag{9}$$

Let $\mathbf{H}_{(norm)} = (\|\mathbf{H}_1\|_2, \|\mathbf{H}_2\|_2, \ldots, \|\mathbf{H}_n\|_2)$. Further, let $|\mathbf{H}_{(norm)}|_{(i)}$ be the $i$-th smallest of the elements of $\mathbf{H}_{(norm)}$. Set $\tilde{\mathbf{H}} = (|\mathbf{H}_{(norm)}|_{(1)}, |\mathbf{H}_{(norm)}|_{(2)}, \ldots, |\mathbf{H}_{(norm)}|_{(n)})^T$. If $\mathbf{w} \in S_s$ then a vector obtained by permuting the blocks of $\mathbf{w}$ in any possible way is also in $S_s$. Then (9) can be rewritten as

$$w(\mathbf{h}, S_s) = \max_{\mathbf{w} \in S_s} \sum_{i=1}^{n} \tilde{\mathbf{H}}_i \|\mathbf{W}_i\|_2 \tag{10}$$

where $\tilde{\mathbf{H}}_i$ is the $i$-th element of vector $\tilde{\mathbf{H}}$. Let $\hat{\mathbf{w}}$ be the solution of the maximization on the right-hand side of (10). Further let $\hat{\mathbf{W}}_i = (\hat{\mathbf{w}}_{(i-1)d+1}, \hat{\mathbf{w}}_{(i-1)d+2}, \ldots, \hat{\mathbf{w}}_{id})^T, i = 1, 2, \ldots, n$. It then easily follows $\|\hat{\mathbf{W}}_n\|_2 \geq \|\hat{\mathbf{W}}_{n-1}\|_2 \geq \|\hat{\mathbf{W}}_{n-2}\|_2 \geq \cdots \geq \|\hat{\mathbf{W}}_1\|_2$. To see this assume that there is a pair of indexes $n_1, n_2$ such that $n_1 < n_2$ and $\|\hat{\mathbf{W}}_{n_1}\|_2 > \|\hat{\mathbf{W}}_{n_2}\|_2$. However, $\|\hat{\mathbf{W}}_{n_1}\|_2 \tilde{\mathbf{H}}_{n_1} + \|\hat{\mathbf{W}}_{n_2}\|_2 \tilde{\mathbf{H}}_{n_2} < \|\hat{\mathbf{W}}_{n_2}\|_2 \tilde{\mathbf{H}}_{n_1} + \|\hat{\mathbf{W}}_{n_1}\|_2 \tilde{\mathbf{H}}_{n_2}$ and $\hat{\mathbf{w}}$ would not be the optimal solution of the maximization on the right-hand side of (10).

Let $\mathbf{y} = (\mathbf{y}_1, \mathbf{y}_2, \ldots, \mathbf{y}_n)^T \in R^n$. Then one can simplify (10) in the following way

$$\begin{aligned}
w(\mathbf{h}, S_s) = \max_{\mathbf{y} \in R^n} \quad & \sum_{i=1}^{n} \tilde{\mathbf{H}}_i \mathbf{y}_i \\
\text{subject to} \quad & \mathbf{y}_i \geq 0, 0 \leq i \leq n \\
& \sum_{i=n-k+1}^{n} \mathbf{y}_i \geq \sum_{i=1}^{n-k} \mathbf{y}_i \\
& \sum_{i=1}^{n} \mathbf{y}_i^2 \leq 1.
\end{aligned} \tag{11}$$

One can add the sorting constraints on the elements of $\mathbf{y}$ in the optimization problem above. However, they would be redundant, i.e. any solution $\hat{\mathbf{y}}$ to the above optimization problem will automatically satisfy $\hat{\mathbf{y}}_n \geq \hat{\mathbf{y}}_{n-1} \geq \cdots \geq \hat{\mathbf{y}}_1$. To determine an upper bound on $w(\mathbf{h}, S_s)$ we will use the method of Lagrange duality. The derivation of Lagrange dual upper bound will closely follow a similar derivation from [82]. For the completeness we reproduce it here as well. Before deriving the Lagrange dual we slightly modify (11) in the following way

$$-w(\mathbf{h}, S_s) = \min_{\mathbf{y} \in R^n} \quad -\sum_{i=1}^{n} \tilde{\mathbf{H}}_i \mathbf{y}_i$$
$$\text{subject to} \quad \mathbf{y}_i \geq 0, 0 \leq i \leq n$$
$$\sum_{i=n-k+1}^{n} \mathbf{y}_i \geq \sum_{i=1}^{n-k} \mathbf{y}_i$$
$$\sum_{i=1}^{n} \mathbf{y}_i^2 \leq 1. \quad (12)$$

To further facilitate writing let $\mathbf{z} \in R^n$ be a column vector such that $\mathbf{z}_i = 1, 1 \leq i \leq (n-k)$ and $\mathbf{z}_i = -1, n-k+1 \leq i \leq n$. Further, let $\lambda = (\lambda_1, \lambda_2, \ldots, \lambda_n)^T \in R^n$. Following, e.g. [11], we can write the dual of the optimization problem (12) and its optimal value $w_{up}(\mathbf{h}, S_s)$ as

$$-w_{up}(\mathbf{h}, S_s) = \max_{\gamma, \nu, \lambda} \min_{\mathbf{y}} \quad -\tilde{\mathbf{H}}^T \mathbf{y} + \gamma \|\mathbf{y}\|_2^2 - \gamma + \nu \mathbf{z}^T \mathbf{y} - \lambda^T \mathbf{y}$$
$$\text{subject to} \quad \nu \geq 0, \gamma \geq 0$$
$$\lambda_i \geq 0, 0 \leq i \leq n. \quad (13)$$

One can then transform the objective function in the following way

$$-w_{up}(\mathbf{h}, S_s) = \max_{\gamma, \nu, \lambda} \min_{\mathbf{y}} \quad \|\sqrt{\gamma}\mathbf{y} - \frac{\lambda + \tilde{\mathbf{H}} - \nu \mathbf{z}}{2\sqrt{\gamma}}\|_2^2 - \gamma - \frac{\|\lambda + \tilde{\mathbf{H}} - \nu \mathbf{z}\|_2^2}{4\gamma}$$
$$\text{subject to} \quad \nu \geq 0, \gamma \geq 0$$
$$\lambda_i \geq 0, 0 \leq i \leq n. \quad (14)$$

After trivially solving the inner minimization in (14) we obtain

$$w_{up}(\mathbf{h}, S_s) = \min_{\gamma, \nu, \lambda} \quad \gamma + \frac{\|\lambda + \tilde{\mathbf{H}} - \nu \mathbf{z}\|_2^2}{4\gamma}$$
$$\text{subject to} \quad \nu \geq 0, \gamma \geq 0$$
$$\lambda_i \geq 0, 0 \leq i \leq n. \quad (15)$$

Minimization over $\gamma$ is straightforward and one easily obtains that $\gamma = \frac{\|\lambda + \tilde{\mathbf{H}} - \nu\mathbf{z}\|_2}{2}$ is optimal. Plugging this value of $\gamma$ back in the objective function of the optimization problem (15) one obtains

$$w_{up}(\mathbf{h}, S_s) = \min_{\nu, \lambda} \quad \|\lambda + \tilde{\mathbf{H}} - \nu\mathbf{z}\|_2$$
$$\text{subject to} \quad \nu \geq 0$$
$$\lambda_i \geq 0, 0 \leq i \leq n. \quad (16)$$

By duality, $-w_{up}(\mathbf{h}, S_s) \leq -w(\mathbf{h}, S_s)$ which easily implies $w(\mathbf{h}, S_s) \leq w_{up}(\mathbf{h}, S_s)$. Therefore $w_{up}(\mathbf{h}, S_s)$ is an upper bound on $w(\mathbf{h}, S_s)$. (In fact one can easily show that the strong duality holds and that $w(\mathbf{h}, S_s) = w_{up}(\mathbf{h}, S_s)$; however, as explained earlier, for our analysis showing that $w_{up}(\mathbf{h}, S_s)$ is an upper bound on $w(\mathbf{h}, S_s)$ is sufficient.) Along the same lines, one can easily spot that any feasible values $\nu$ and $\lambda$ in (16) will provide a valid upper bound on $w_{up}(\mathbf{h}, S_s)$ and hence a valid upper bound on $w(\mathbf{h}, S_s)$. In what follows we will in fact determine the optimal values for $\nu$ and $\lambda$. However, since it is not necessary for our analysis we will not put too much effort into proving that these values are optimal. As we have stated earlier, for our analysis it will be enough to show that the values for $\nu$ and $\lambda$ that we will obtain are feasible in (16).

To facilitate the exposition in what follows instead of dealing with the objective function given in (16) we will be dealing with its squared value. Hence, we set $f(\mathbf{h}, \nu, \lambda) = \|\lambda + \tilde{\mathbf{H}} - \nu\mathbf{z}\|_2^2$. Now, let $\lambda = (\lambda_1, \lambda_2, \ldots, \lambda_c, 0, 0, \ldots, 0)^T, \lambda_1 \geq \lambda_2 \geq \cdots \geq \lambda_c \geq 0$ where $c \leq (n-k)$ is a crucial parameter that will be determined later. The optimization over $\nu$ in (16) is then seemingly straightforward. Setting the derivative of $f(\mathbf{h}, \nu, \lambda)$ with respect to $\nu$ to zero we have

$$\frac{d\|\lambda + \tilde{\mathbf{H}} - \nu\mathbf{z}\|_2^2}{d\nu} = 0$$
$$\Leftrightarrow -2(\lambda + \tilde{\mathbf{H}})^T \mathbf{z} + 2\|\mathbf{z}\|_2^2 \nu = 0$$
$$\Leftrightarrow \nu = \frac{(\lambda + \tilde{\mathbf{H}})^T \mathbf{z}}{\|\mathbf{z}\|_2^2}. \quad (17)$$

If $(\lambda + \tilde{\mathbf{H}})^T \mathbf{z} \geq 0$ then $\nu = \frac{(\lambda + \tilde{\mathbf{H}})^T \mathbf{z}}{\|\mathbf{z}\|_2^2}$ is indeed the optimal in (16). For the time being let us assume that $\lambda, \mathbf{h}, c$ are such that $\nu = \frac{(\lambda + \tilde{\mathbf{H}})^T \mathbf{z}}{\|\mathbf{z}\|_2^2} \geq 0$. For $\nu = \frac{(\lambda + \tilde{\mathbf{H}})^T \mathbf{z}}{\|\mathbf{z}\|_2^2}$ we have

$$f(\mathbf{h}, \frac{(\lambda + \tilde{\mathbf{H}})^T \mathbf{z}}{\|\mathbf{z}\|_2^2}, \lambda) = \|(\lambda + \tilde{\mathbf{H}})^T (I - \frac{\mathbf{z}\mathbf{z}^T}{\mathbf{z}^T \mathbf{z}})\|_2^2 = (\lambda + \tilde{\mathbf{H}})^T (I - \frac{\mathbf{z}\mathbf{z}^T}{\mathbf{z}^T \mathbf{z}})(\lambda + \tilde{\mathbf{H}}). \quad (18)$$

Simplifying (18) further we obtain

$$f(\mathbf{h}, \frac{(\lambda+\tilde{\mathbf{H}})^T\mathbf{z}}{\|\mathbf{z}\|_2^2}, \lambda) = \sum_{i=1}^{n}\tilde{\mathbf{H}}_i^2 + 2\sum_{i=1}^{c}\lambda_i\tilde{\mathbf{H}}_i + \sum_{i=1}^{c}\lambda_i^2 - \frac{(\tilde{\mathbf{H}}^T\mathbf{z})^2}{n} - \frac{(\sum_{i=1}^{c}\lambda_i)^2}{n} - \frac{2(\sum_{i=1}^{c}\lambda_i)(\tilde{\mathbf{H}}^T\mathbf{z})}{n}. \quad (19)$$

To determine good values for $\lambda$ we proceed by setting the derivatives of $f(\mathbf{h}, \frac{(\lambda+\tilde{\mathbf{H}})^T\mathbf{z}}{\|\mathbf{z}\|_2^2}, \lambda)$ with respect to $\lambda_i, 1 \leq i \leq c$ to zero

$$\frac{df(\mathbf{h}, \frac{(\lambda+\tilde{\mathbf{H}})^T\mathbf{z}}{\|\mathbf{z}\|_2^2}, \lambda)}{d\lambda_i} = 2\lambda_i + 2\tilde{\mathbf{H}}_i - 2\frac{(\sum_{i=1}^{c}\lambda_i)}{n} - 2\frac{(\tilde{\mathbf{H}}^T\mathbf{z})}{n} = 0. \quad (20)$$

Summing the above derivatives over $i$ and equalling with zero we obtain

$$\sum_{i=1}^{c}\frac{df(\mathbf{h}, \frac{(\lambda+\tilde{\mathbf{H}})^T\mathbf{z}}{\|\mathbf{z}\|_2^2}, \lambda)}{d\lambda_i} = 2(\sum_{i=1}^{c}\lambda_i + \sum_{i=1}^{c}\tilde{\mathbf{H}}_i - c\frac{(\sum_{i=1}^{c}\lambda_i)}{n} - c\frac{(\tilde{\mathbf{H}}^T\mathbf{z})}{n}) = 0. \quad (21)$$

From (21) one then easily finds

$$\sum_{i=1}^{c}\lambda_i = \frac{c(\tilde{\mathbf{H}}^T\mathbf{z})}{n-c} - \frac{n\sum_{i=1}^{c}\tilde{\mathbf{H}}_i}{n-c}. \quad (22)$$

Plugging the value for $\sum_{i=1}^{c}\lambda_i$ obtained in (22) in (20) we have

$$\lambda_i = \frac{(\tilde{\mathbf{H}}^T\mathbf{z})}{n} - \tilde{\mathbf{H}}_i + \frac{(\sum_{i=1}^{c}\lambda_i)}{n} = \frac{(\tilde{\mathbf{H}}^T\mathbf{z})}{n} - \tilde{\mathbf{H}}_i + \frac{c(\tilde{\mathbf{H}}^T\mathbf{z})}{n(n-c)} - \frac{\sum_{i=1}^{c}\tilde{\mathbf{H}}_i}{n-c}$$

and finally

$$\begin{aligned}\lambda_i &= \frac{(\tilde{\mathbf{H}}^T\mathbf{z}) - \sum_{i=1}^{c}\tilde{\mathbf{H}}_i}{n-c} - \tilde{\mathbf{H}}_i, 1 \leq i \leq c \\ \lambda_i &= 0, c+1 \leq i \leq n.\end{aligned} \quad (23)$$

Combining (17) and (22) we have

$$\nu = \frac{(\lambda+\tilde{\mathbf{H}})^T\mathbf{z}}{\|\mathbf{z}\|_2^2} = \frac{\tilde{\mathbf{H}}^T\mathbf{z} + \sum_{i=1}^{c}\lambda_i}{n} = \frac{\tilde{\mathbf{H}}^T\mathbf{z} + \frac{c(\tilde{\mathbf{H}}^T\mathbf{z})}{n-c} - \frac{n\sum_{i=1}^{c}\tilde{\mathbf{H}}_i}{n-c}}{n} = \frac{(\tilde{\mathbf{H}}^T\mathbf{z}) - \sum_{i=1}^{c}\tilde{\mathbf{H}}_i}{n-c}. \quad (24)$$

From (23) we then have as expected

$$\nu = \lambda_i + \tilde{\mathbf{H}}_i, 1 \leq i \leq c. \quad (25)$$

As long as we can find a $c$ such that $\lambda_i, 1 \leq i \leq c$ given in (23) are non-negative $\nu$ will be non-negative as well and $\nu$ and $\lambda$ will therefore be feasible in (16). This in turn implies

$$w(\mathbf{h}, S_s) \leq \sqrt{f(\mathbf{h}, \nu, \lambda)} \tag{26}$$

where $f(\mathbf{h}, \nu, \lambda)$ is computed for the values of $\lambda$ and $\nu$ given in (23) and (25), respectively. (In fact determining the largest $c$ such that $\lambda_i, 1 \leq i \leq c$ given in (23) are non-negative will insure that $\sqrt{f(\mathbf{h}, \nu, \lambda)} = w(\mathbf{h}, S_s)$; however, as already stated earlier, this fact is not of any special importance for our analysis).

Let us now assume that $c$ is fixed such that $\lambda$ and $\nu$ are as given in (23) and (25). Then combining (19), (22), and (25) we have

$$f(\mathbf{h}, \frac{(\lambda + \tilde{\mathbf{H}})^T \mathbf{z}}{\|\mathbf{z}\|_2^2}, \lambda) = \sum_{i=1}^n \tilde{\mathbf{H}}_i^2 + 2\nu \sum_{i=1}^c \tilde{\mathbf{H}}_i - 2 \sum_{i=1}^c \tilde{\mathbf{H}}_i^2 + c\nu^2 - 2\nu \sum_{i=1}^c \tilde{\mathbf{H}}_i + \sum_{i=1}^c \tilde{\mathbf{H}}_i^2 - \frac{(\sum_{i=1}^c \lambda_i + \tilde{\mathbf{H}}^T \mathbf{z})^2}{n}. \tag{27}$$

Combining (22) and (24) we obtain

$$(\sum_{i=1}^c \lambda_i + \tilde{\mathbf{H}}^T \mathbf{z}) = n\nu. \tag{28}$$

Further, combining (27) and (28) we find

$$\begin{aligned} f(\mathbf{h}, \frac{(\lambda + \tilde{\mathbf{H}})^T \mathbf{z}}{\|\mathbf{z}\|_2^2}, \lambda) &= \sum_{i=1}^n \tilde{\mathbf{H}}_i^2 + c\nu^2 - \sum_{i=1}^c \tilde{\mathbf{H}}_i^2 - \frac{(n\nu)^2}{n} \\ &= \sum_{i=1}^n \tilde{\mathbf{H}}_i^2 + (c-n)\nu^2 - \sum_{i=1}^c \tilde{\mathbf{H}}_i^2 \\ &= \sum_{i=1}^n \tilde{\mathbf{H}}_i^2 - \sum_{i=1}^c \tilde{\mathbf{H}}_i^2 - \frac{((\tilde{\mathbf{H}}^T \mathbf{z}) - \sum_{i=1}^c \tilde{\mathbf{H}}_i)^2}{n-c}. \end{aligned} \tag{29}$$

Finally, combining (26) and (29) we have

$$w(\mathbf{h}, S_s) \leq \sqrt{\sum_{i=1}^n \tilde{\mathbf{H}}_i^2 - \sum_{i=1}^c \tilde{\mathbf{H}}_i^2 - \frac{((\tilde{\mathbf{H}}^T \mathbf{z}) - \sum_{i=1}^c \tilde{\mathbf{H}}_i)^2}{n-c}} = \sqrt{\sum_{i=c+1}^n \tilde{\mathbf{H}}_i^2 - \frac{((\tilde{\mathbf{H}}^T \mathbf{z}) - \sum_{i=1}^c \tilde{\mathbf{H}}_i)^2}{n-c}}. \tag{30}$$

Clearly, as long as $(\tilde{\mathbf{H}}^T \mathbf{z}) \geq 0$ there will be a $c \leq n-k$ (it is possible that $c = 0$) such that quantity on the most right hand side of (30) is an upper bound on $w(\mathbf{h}, S_s)$.

To facilitate the exposition in the following subsection we will make the upper bound given in (30) slightly more pessimistic in the following lemma.

**Lemma 1.** *Let $\mathbf{h} \in R^{dn}$ be a vector with i.i.d. zero-mean unit variance gaussian components. Let $\mathbf{H}_i = (\mathbf{h}_{(i-1)d+1}, \mathbf{h}_{(i-1)d+2}, \ldots, \mathbf{h}_{id})^T$, $i = 1, 2, \ldots, n$ and $\mathbf{H}_{(norm)} = (\|\mathbf{H}_1\|_2, \|\mathbf{H}_2\|_2, \ldots, \|\mathbf{H}_n\|_2)$. Further,* let $|\mathbf{H}_{(norm)}|_{(i)}$ be the $i$-th smallest of the elements of $\mathbf{H}_{(norm)}$. Set $\tilde{\mathbf{H}} = (|\mathbf{H}_{(norm)}|_{(1)}, |\mathbf{H}_{(norm)}|_{(2)}, \ldots,$
$|\mathbf{H}_{(norm)}|_{(n)})^T$ and $w(\mathbf{h}, S_s) = \max_{\mathbf{w} \in S_s}(\mathbf{h}^T \mathbf{w})$ where $S_s$ is as defined in (7). Let $\mathbf{z} \in R^n$ be a column vector such that $\mathbf{z}_i = 1, 1 \leq i \leq (n-k)$ and $\mathbf{z}_i = -1, n-k+1 \leq i \leq n$. Then

$$w(\mathbf{h}, S_s) \leq B_s \tag{31}$$

where

$$B_s = \begin{cases} \sqrt{\sum_{i=1}^n \tilde{\mathbf{H}}_i^2} & \text{if} \quad \zeta_s(\mathbf{h}, c_s) \leq 0 \\ \sqrt{\sum_{i=c_s+1}^n \tilde{\mathbf{H}}_i^2 - \frac{((\tilde{\mathbf{H}}^T \mathbf{z}) - \sum_{i=1}^{c_s} \tilde{\mathbf{H}}_i)^2}{n - c_s}} & \text{if} \quad \zeta_s(\mathbf{h}, c_s) > 0 \end{cases}, \tag{32}$$

$\zeta_s(\mathbf{h}, c) = \frac{(\tilde{\mathbf{H}}^T \mathbf{z}) - \sum_{i=1}^c \tilde{\mathbf{H}}_i}{n-c} - \tilde{\mathbf{H}}_c$ and $c_s = \delta_s n$ is a $c \leq n-k$ such that

$$\frac{(1-\epsilon)E((\tilde{\mathbf{H}}^T \mathbf{z}) - \sum_{i=1}^c \tilde{\mathbf{H}}_i)}{n-c} - F_{\chi_d}^{-1}\left(\frac{(1+\epsilon)c}{n}\right) = 0. \tag{33}$$

$F_{\chi_d}^{-1}(\cdot)$ is the inverse cdf of the chi random variable with $d$ degrees of freedom, i.e. it is the inverse cdf of random variable $\sqrt{\sum_{i=1}^d Z_i^2}$ where $Z_i, 1 \leq i \leq d$ are independent zero-mean, unit variance Gaussian random variables. $\epsilon > 0$ is an arbitrarily small constant independent of $n$.

*Proof.* Follows from the previous analysis and (30). □

### 4.1.2 Computing an upper bound on $E(B_s)$

In this subsection we will compute an upper bound on $E(B_s)$. Again, the derivation will closely follow that of [82]. (However, due to a few block-structure related differences in the derivations of Lemmas 2 and 3 we include it here.) As a first step we determine a lower bound on $P(\zeta_s(\mathbf{h}, c_s) > 0)$. We start by a sequence of obvious inequalities

$$\begin{aligned} P(\zeta_s(\mathbf{h}, c_s) > 0) &\geq P\left(\zeta_s(\mathbf{h}, c_s) \geq \frac{(1-\epsilon)E((\tilde{\mathbf{H}}^T \mathbf{z}) - \sum_{i=1}^{c_s} \tilde{\mathbf{H}}_i)}{n - c_s} - F_{\chi_d}^{-1}\left(\frac{(1+\epsilon)c_s}{n}\right)\right) \\ &\geq P\left(\frac{((\tilde{\mathbf{H}}^T \mathbf{z}) - \sum_{i=1}^{c_s} \tilde{\mathbf{H}}_i)}{n - c_s} \geq \frac{(1-\epsilon)E((\tilde{\mathbf{H}}^T \mathbf{z}) - \sum_{i=1}^{c_s} \tilde{\mathbf{H}}_i)}{n - c_s} \text{ and } F_{\chi_d}^{-1}\left(\frac{(1+\epsilon)c_s}{n}\right) \geq \tilde{\mathbf{H}}_{c_s}\right) \\ &\geq 1 - P\left(\frac{((\tilde{\mathbf{H}}^T \mathbf{z}) - \sum_{i=1}^{c_s} \tilde{\mathbf{H}}_i)}{n - c_s} < \frac{(1-\epsilon)E((\tilde{\mathbf{H}}^T \mathbf{z}) - \sum_{i=1}^{c_s} \tilde{\mathbf{H}}_i)}{n - c_s}\right) - P\left(F_{\chi_d}^{-1}\left(\frac{(1+\epsilon)c_s}{n}\right) < \tilde{\mathbf{H}}_{c_s}\right) \end{aligned} \tag{34}$$

The rest of the analysis assumes that $n$ is large so that $\delta_s$ can be assumed to be real (of course, $\delta_s$ is a proportionality constant independent of $n$). Using the results from [7] we obtain

$$P\left(F_{\chi_d}^{-1}\left(\frac{(1+\epsilon)c_s}{n}\right) < \tilde{\mathbf{H}}_{c_s}\right) \leq \exp\left\{-\frac{n}{2\frac{(1+\epsilon)c_s}{n}}\left(\frac{c_s}{n} - \frac{(1+\epsilon)c_s}{n}\right)^2\right\}$$
$$\leq \exp\left\{-\frac{n\epsilon^2 \delta_s}{2(1+\epsilon)}\right\}. \tag{35}$$

We will also need the following brilliant result from [20]. Let $\xi(\cdot) : R^{dn} \longrightarrow R$ be a Lipschitz function such that $|\xi(\mathbf{a}) - \xi(\mathbf{b})| \leq \sigma \|\mathbf{a} - \mathbf{b}\|_2$. Let $\mathbf{a}$ be a vector comprised of i.i.d. zero-mean, unit variance Gaussian random variables. Then

$$P((1-\epsilon)E\xi(\mathbf{a}) \geq \xi(\mathbf{a})) \leq \exp\left\{-\frac{(\epsilon E\xi(\mathbf{a}))^2}{2\sigma^2}\right\}. \tag{36}$$

Let $\xi(\mathbf{h}) = (\tilde{\mathbf{H}}^T \mathbf{z}) - \sum_{i=1}^{c_s} \tilde{\mathbf{H}}_i$. The following lemma estimates $\sigma$ (for simplicity we assume $c_s = 0$; the proof easily extends to the case when $c_s \neq 0$).

**Lemma 2.** *Let $\mathbf{a}, \mathbf{b} \in R^{dn}$. Let $\mathbf{A}_i = (\mathbf{a}_{(i-1)d+1}, \mathbf{a}_{(i-1)d+2}, \ldots, \mathbf{a}_{id})$ and $\mathbf{B}_i = (\mathbf{b}_{(i-1)d+1}, \mathbf{b}_{(i-1)d+2}, \ldots, \mathbf{b}_{id})$, $i = 1, 2, \ldots, n$. Set $\mathbf{A}_{(norm)} = (\|\mathbf{A}_1\|_2, \|\mathbf{A}_2\|_2, \ldots, \|\mathbf{A}_n\|_2)$ and $\mathbf{B}_{(norm)} = (\|\mathbf{B}_1\|_2, \|\mathbf{B}_2\|_2, \ldots, \|\mathbf{B}_n\|_2)$. Further, let $|\mathbf{A}_{(norm)}|_{(i)}, |\mathbf{B}_{(norm)}|_{(i)}$ be the $i$-th smallest of the elements of $\mathbf{A}_{(norm)}, \mathbf{B}_{(norm)}$, respectively. Set $\tilde{\mathbf{A}} = (|\mathbf{A}_{(norm)}|_{(1)}, |\mathbf{A}_{(norm)}|_{(2)}, \ldots, |\mathbf{A}_{(norm)}|_{(n)})^T$ and $\tilde{\mathbf{B}} = (|\mathbf{B}_{(norm)}|_{(1)}, |\mathbf{B}_{(norm)}|_{(2)}, \ldots, |\mathbf{B}_{(norm)}|_{(n)})^T$. Then*

$$|\xi(\mathbf{a}) - \xi(\mathbf{b})| = |\sum_{i=1}^{n-k} \tilde{\mathbf{A}}_i - \sum_{i=n-k+1}^{n} \tilde{\mathbf{A}}_i - \sum_{i=1}^{n-k} \tilde{\mathbf{B}}_i + \sum_{i=n-k+1}^{n} \tilde{\mathbf{B}}_i| \leq \sqrt{n}\sqrt{\sum_{i=1}^{dn} |\mathbf{a}_i - \mathbf{b}_i|^2} = \sqrt{n}\|\mathbf{a} - \mathbf{b}\|_2. \tag{37}$$

*Proof.* We have

$$|\sum_{i=1}^{n-k} \tilde{\mathbf{A}}_i - \sum_{i=n-k+1}^{n} \tilde{\mathbf{A}}_i - \sum_{i=1}^{n-k} \tilde{\mathbf{B}}_i + \sum_{i=n-k+1}^{n} \tilde{\mathbf{B}}_i| \leq |\sum_{i=1}^{n-k}(\tilde{\mathbf{A}}_i - \tilde{\mathbf{B}}_i)| + |\sum_{i=n-k+1}^{n}(\tilde{\mathbf{A}}_i - \tilde{\mathbf{B}}_i)|$$
$$\leq \sum_{i=1}^{n-k} |\tilde{\mathbf{A}}_i - \tilde{\mathbf{B}}_i| + \sum_{i=n-k+1}^{n} |\tilde{\mathbf{A}}_i - \tilde{\mathbf{B}}_i| \leq \sum_{i=1}^{n} |\tilde{\mathbf{A}}_i - \tilde{\mathbf{B}}_i| \leq \sqrt{n}\sqrt{\sum_{i=1}^{n} |\tilde{\mathbf{A}}_i - \tilde{\mathbf{B}}_i|^2}$$
$$\leq \sqrt{n}\sqrt{\sum_{i=1}^{n} |\tilde{\mathbf{A}}_i|^2 + \sum_{i=1}^{n} |\tilde{\mathbf{B}}_i|^2 - 2\sum_{i=1}^{n} \tilde{\mathbf{A}}_i \tilde{\mathbf{B}}_i} = \sqrt{n}\sqrt{\sum_{i=1}^{dn} |\mathbf{a}_i|^2 + \sum_{i=1}^{dn} |\mathbf{b}_i|^2 - 2\sum_{i=1}^{n} \tilde{\mathbf{A}}_i \tilde{\mathbf{B}}_i}. \tag{38}$$

Since the components of $\tilde{\mathbf{A}}$ and $\tilde{\mathbf{B}}$ are positive and sorted in the same non-decreasing order we have

$$\sum_{i=1}^{n} \tilde{\mathbf{A}}_i \tilde{\mathbf{B}}_i \geq \sum_{i=1}^{n} \|\mathbf{A}_i\|_2 \|\mathbf{B}_i\|_2. \tag{39}$$

By Cauchy-Schwartz inequality we have

$$\sum_{i=1}^{n} \|\mathbf{A}_i\|_2 \|\mathbf{B}_i\|_2 \geq \sum_{i=1}^{n} \sum_{j=1}^{d} \mathbf{a}_{(i-1)d+j} \mathbf{b}_{(i-1)d+j} = \sum_{i=1}^{dn} \mathbf{a}_i \mathbf{b}_i. \tag{40}$$

From (39) and (40) we obtain

$$-\sum_{i=1}^{n} \tilde{\mathbf{A}}_i \tilde{\mathbf{B}}_i \leq -\sum_{i=1}^{dn} \mathbf{a}_i \mathbf{b}_i. \tag{41}$$

Combining (38) and (41) we finally have

$$|\sum_{i=1}^{n-k} \tilde{\mathbf{A}}_i - \sum_{i=n-k+1}^{n} \tilde{\mathbf{A}}_i - \sum_{i=1}^{n-k} \tilde{\mathbf{B}}_i + \sum_{i=n-k+1}^{n} \tilde{\mathbf{B}}_i| \leq \sqrt{n} \sqrt{\sum_{i=1}^{dn} |\mathbf{a}_i|^2 + \sum_{i=1}^{dn} |\mathbf{b}_i|^2 - 2\sum_{i=1}^{n} \tilde{\mathbf{A}}_i \tilde{\mathbf{B}}_i}$$

$$\leq \sqrt{n} \sqrt{\sum_{i=1}^{dn} |\mathbf{a}_i|^2 + \sum_{i=1}^{dn} |\mathbf{b}_i|^2 - 2\sum_{i=1}^{dn} \mathbf{a}_i \mathbf{b}_i} = \sqrt{n} \sqrt{\sum_{i=1}^{dn} |\mathbf{a}_i - \mathbf{b}_i|^2}. \tag{42}$$

Connecting beginning and end in (42) establishes (37). □

For $\xi(\mathbf{h}) = (\tilde{\mathbf{H}}^T \mathbf{z}) - \sum_{i=1}^{c_s} \tilde{\mathbf{H}}_i$ the previous lemma then gives $\sigma \leq \sqrt{n}$ (in fact if there was no assumption that $c_s = 0$ one would rather handily obtain $\sigma \leq \sqrt{n - c_s}$ by merely recognizing that the length of all relevant vectors would be $\sigma \leq \sqrt{n - c_s}$ instead of $n$). As shown in [77] (and as we will see later in this paper), if $n$ is large and $\delta_s$ is a constant independent of $n$, $E((\tilde{\mathbf{H}}^T \mathbf{z}) - \sum_{i=1}^{c_s} \tilde{\mathbf{H}}_i) = \psi_s n$ where $\psi_s$ is independent of $n$ as well ($\psi_s$ is of course dependent on $\beta$ and $\delta_s$). Hence (36) with $\xi(\mathbf{h}) = (\tilde{\mathbf{H}}^T \mathbf{z}) - \sum_{i=1}^{c_s} \tilde{\mathbf{H}}_i$ gives us

$$P\left(\frac{((\tilde{\mathbf{H}}^T \mathbf{z}) - \sum_{i=1}^{c_s} \tilde{\mathbf{H}}_i)}{n - c_s} < \frac{(1-\epsilon) E((\tilde{\mathbf{H}}^T \mathbf{z}) - \sum_{i=1}^{c_s} \tilde{\mathbf{H}}_i)}{n - c_s}\right) \leq \exp\left\{-\frac{(\epsilon \psi_s n)^2}{2n}\right\} = \exp\left\{-\frac{\epsilon^2 \psi_s^2 n}{2}\right\}. \tag{43}$$

Combining (34), (35), and (43) we finally obtain

$$P(\zeta_s(\mathbf{h},c_s)>0) \geq 1 - P\left(\frac{((\tilde{\mathbf{H}}^T\mathbf{z}) - \sum_{i=1}^{c_s}\tilde{\mathbf{H}}_i)}{n-c_s} < \frac{(1-\epsilon)E((\tilde{\mathbf{H}}^T\mathbf{z}) - \sum_{i=1}^{c_s}\tilde{\mathbf{H}}_i)}{n-c_s}\right)$$
$$- P\left(F_{\chi_d}^{-1}\left(\frac{(1+\epsilon)c_s}{n}\right) < \tilde{\mathbf{H}}_{c_s}\right)$$
$$\geq 1 - \exp\left\{-\frac{n\epsilon^2\delta_s}{2(1+\epsilon)}\right\} - \exp\left\{-\frac{\epsilon^2\psi_s^2 n}{2}\right\}. \quad (44)$$

We now return to computing an upper bound on $E(B_s)$. By the definition of $B_s$ we have

$$E(B_s) = \int_{\zeta_s(\mathbf{h},c_s)\leq 0} \sqrt{\sum_{i=1}^{n}\tilde{\mathbf{H}}_i^2}\, p(\mathbf{h})d\mathbf{h} + \int_{\zeta_s(\mathbf{h},c_s)>0} \sqrt{\sum_{i=c_s+1}^{n}\tilde{\mathbf{H}}_i^2 - \frac{((\tilde{\mathbf{H}}^T\mathbf{z}) - \sum_{i=1}^{c_s}\tilde{\mathbf{H}}_i)^2}{n-c_s}}\, p(\mathbf{h})d\mathbf{h} \quad (45)$$

where $p(\mathbf{h})$ is the joint pdf of the i.i.d. zero-mean unit variance gaussian components of vector $\mathbf{h}$. Since the functions $\sqrt{\sum_{i=1}^{n}\tilde{\mathbf{H}}_i^2}$ and $p(\mathbf{h})$ are rotationally invariant and since the region $\zeta_s(\mathbf{h},c_s) \leq 0$ takes up the same fraction of the surface area of sphere of any radius we have

$$\int_{\zeta_s(\mathbf{h},c_s)\leq 0} \sqrt{\sum_{i=1}^{n}\tilde{\mathbf{H}}_i^2}\, p(\mathbf{h})d\mathbf{h} = E\sqrt{\sum_{i=1}^{n}\tilde{\mathbf{H}}_i^2} \int_{\zeta_s(\mathbf{h},c_s)\leq 0} p(\mathbf{h})d\mathbf{h} \leq \sqrt{E\sum_{i=1}^{n}\tilde{\mathbf{H}}_i^2} \int_{\zeta_s(\mathbf{h},c_s)\leq 0} p(\mathbf{h})d\mathbf{h}. \quad (46)$$

Combining (44) and (46) we further have

$$\int_{\zeta_s(\mathbf{h},c_s)\leq 0} \sqrt{\sum_{i=1}^{n}\tilde{\mathbf{H}}_i^2}\, p(\mathbf{h})d\mathbf{h} \leq \sqrt{E\sum_{i=1}^{n}\tilde{\mathbf{H}}_i^2}\left(\exp\left\{-\frac{n\epsilon^2\delta_s}{2(1+\epsilon)}\right\} + \exp\left\{-\frac{\epsilon^2\psi_s^2 n}{2}\right\}\right). \quad (47)$$

It also easily follows

$$\int_{\zeta_s(\mathbf{h},c_s)>0} \sqrt{\sum_{i=c_s+1}^{n}\tilde{\mathbf{H}}_i^2 - \frac{((\tilde{\mathbf{H}}^T\mathbf{z}) - \sum_{i=1}^{c_s}\tilde{\mathbf{H}}_i)^2}{n-c_s}}\, p(\mathbf{h})d\mathbf{h} \leq \int_{\mathbf{h}} \sqrt{\sum_{i=c_s+1}^{n}\tilde{\mathbf{H}}_i^2 - \frac{((\tilde{\mathbf{H}}^T\mathbf{z}) - \sum_{i=1}^{c_s}\tilde{\mathbf{H}}_i)^2}{n-c_s}}\, p(\mathbf{h})d\mathbf{h}$$
$$= E\sqrt{\sum_{i=c_s+1}^{n}\tilde{\mathbf{H}}_i^2 - \frac{((\tilde{\mathbf{H}}^T\mathbf{z}) - \sum_{i=1}^{c_s}\tilde{\mathbf{H}}_i)^2}{n-c_s}} \leq \sqrt{E\sum_{i=c_s+1}^{n}\tilde{\mathbf{H}}_i^2 - \frac{(E(\tilde{\mathbf{H}}^T\mathbf{z}) - E\sum_{i=1}^{c_s}\tilde{\mathbf{H}}_i)^2}{n-c_s}}. \quad (48)$$

Finally, combining (45), (47), and (48) we obtain the following lemma.

**Lemma 3.** *Assume the setup of Lemma 1. Let further $\psi_s = \frac{E((\tilde{\mathbf{H}}^T\mathbf{z}) - \sum_{i=1}^{c_s}\tilde{\mathbf{H}}_i)}{n}$. Then*

$$E(B_s) \leq \sqrt{n}\left(\exp\left\{-\frac{n\epsilon^2\delta_s}{2(1+\epsilon)}\right\} + \exp\left\{-\frac{\epsilon^2\psi_s^2 n}{2}\right\}\right) + \sqrt{E\sum_{i=c_s+1}^{n}\tilde{\mathbf{H}}_i^2 - \frac{(E(\tilde{\mathbf{H}}^T\mathbf{z}) - E\sum_{i=1}^{c_s}\tilde{\mathbf{H}}_i)^2}{n - c_s}}. \tag{49}$$

*Proof.* Follows from the previous discussion. □

If $n$ is large the first term on the right hand side of (49) goes to zero. In a fashion similar to the one presented in [82] from (6), (8), and (49) it then easily follows that for a fixed $\alpha$ one can determine $\beta_s$ as a maximum $\beta$ such that

$$\alpha d > \frac{E\sum_{i=c_s+1}^{n}\tilde{\mathbf{H}}_i^2}{n} - \frac{(E(\tilde{\mathbf{H}}^T\mathbf{z}) - E\sum_{i=1}^{c_s}\tilde{\mathbf{H}}_i)^2}{n(n - c_s)}. \tag{50}$$

As earlier $k = \beta n$ and $\mathbf{z} \in R^n$ is a column vector such that $\mathbf{z}_i = 1, 1 \leq i \leq (n-k)$ and $\mathbf{z}_i = -1, n-k+1 \leq i \leq n$ ($\beta$ is therefore hidden in the above equation in $\mathbf{z}$). As in [82], finding $\beta_s$ for a given fixed $\alpha$ is equivalent to finding minimum $\alpha$ such that (50) holds for a fixed $\beta_s$. Let $\beta_s^{max}$ be $\beta_s$ such that minimum $\alpha$ that satisfies (50) is 1. Our goal is then to determine minimum $\alpha$ that satisfies (50) for any $\beta_s \in [0, \beta_s^{max}]$.

In the rest of this subsection we show how the left hand side of (50) can be computed for a randomly chosen fixed $\beta_s$. As in [82] we do so in two steps:

1. We first determine $c_s$
2. We then compute $\lim_{n\to\infty}\left(\frac{E\sum_{i=c_s+1}^{n}\tilde{\mathbf{H}}_i^2}{n} - \frac{(E(\tilde{\mathbf{H}}^T\mathbf{z}) - E\sum_{i=1}^{c_s}\tilde{\mathbf{H}}_i)^2}{n(n-c_s)}\right)$ with $c_s$ found in step 1.

*Step 1:*

From Lemma 1 we have $c_s = \delta_s n$ is a $c$ such that

$$\frac{(1-\epsilon)E((\sum_{i=1}^{n-\beta_s n}\tilde{\mathbf{H}}_i - \sum_{i=n-\beta_s n+1}^{n}\tilde{\mathbf{H}}_i) - \sum_{i=1}^{c}\tilde{\mathbf{H}}_i)}{n-c} - F_{\chi_d}^{-1}\left(\frac{(1+\epsilon)c}{n}\right) = 0$$

$$\Leftrightarrow \frac{(1-\epsilon)(E\sum_{i=\delta_s n+1}^{n}\tilde{\mathbf{H}}_i - 2E\sum_{i=n-\beta_s n+1}^{n}\tilde{\mathbf{H}}_i)}{n(1-\delta_s)} - F_{\chi_d}^{-1}\left(\frac{(1+\epsilon)\delta_s n}{n}\right) = 0 \tag{51}$$

where as in Lemma 1 $\tilde{\mathbf{H}}_i = |\mathbf{H}_{(norm)}|_{(i)}$ and $|\mathbf{H}_{(norm)}|_{(i)}$ is the $i$-th smallest magnitude of blocks $\mathbf{H}_i$ of $\mathbf{h}$. We also recall that $\mathbf{h} \in R^{dn}$ is a vector with i.i.d. zero-mean unit variance Gaussian random variables and $\epsilon > 0$ is an arbitrarily small constant. Set $\theta_s = 1 - \delta_s$. Following [8, 77] we have

$$\lim_{n\to\infty}\frac{E\sum_{i=(1-\theta_s)n+1}^{n}\tilde{\mathbf{H}}_i}{n} = \int_{F_{\chi_d}^{-1}(1-\theta_s)}^{\infty} t \, dF_{\chi_d}(t), \tag{52}$$

where we recall that $F_{\chi_d}(\cdot)$ is the cdf of any of $\|\mathbf{H}_i\|_2$. Clearly, $\|\mathbf{H}_i\|_2$ is a chi-distributed random variable with $d$ degrees of freedom. We then have for its pdf

$$dF_{\chi_d}(t) = \frac{2^{1-\frac{d}{2}}}{\Gamma(\frac{d}{2})} t^{d-1} e^{-\frac{t^2}{2}}, t \geq 0 \tag{53}$$

where $\Gamma(\cdot)$ stands for the gamma function. The following integration then gives us $F_{\chi_d}^{-1}(1-\theta_s)$. Namely,

$$\frac{2^{1-\frac{k}{2}}}{\Gamma(\frac{d}{2})} \int_0^{F_{\chi_d}^{-1}(1-\theta_s)} t^{d-1} e^{-\frac{t^2}{2}} dt = 1 - \theta_s$$

$$\implies F_{\chi_d}^{-1}(1-\theta_s) = \sqrt{2\gamma_{inc}^{-1}(1-\theta_s, \frac{d}{2})} \tag{54}$$

where $\gamma_{inc}^{-1}(1-\theta_s, \frac{d}{2})$ stands for the inverse of the incomplete gamma function with $\frac{d}{2}$ degrees of freedom evaluated at $(1-\theta_s)$. We further then find

$$\int_{F_{\chi_d}^{-1}(1-\theta_s)}^{\infty} t dF_{\chi_d}(t) = \frac{2^{1-\frac{k}{2}}}{\Gamma(\frac{d}{2})} \int_{F_{\chi_d}^{-1}(1-\theta_s)}^{\infty} t^d e^{-\frac{t^2}{2}} dt = \frac{\sqrt{2}\Gamma(\frac{d+1}{2})}{\Gamma(\frac{d}{2})} \left(1 - \gamma_{inc}(\frac{(F_{\chi_d}^{-1}(1-\theta_s))^2}{2}, \frac{d}{2})\right) \tag{55}$$

where $\gamma_{inc}(\frac{(F_{\chi_d}^{-1}(1-\theta_s))^2}{2}, \frac{d}{2})$ stands for the incomplete gamma function with $\frac{d}{2}$ degrees of freedom evaluated at $\frac{(F_{\chi_d}^{-1}(1-\theta_s))^2}{2}$. From (54) and (55) we obtain

$$\int_{F_{\chi_d}^{-1}(1-\theta_s)}^{\infty} t dF_{\chi_d}(t) = \frac{\sqrt{2}\Gamma(\frac{d+1}{2})}{\Gamma(\frac{d}{2})} \left(1 - \gamma_{inc}(\gamma_{inc}^{-1}(1-\theta_s, \frac{d}{2}), \frac{d+1}{2})\right). \tag{56}$$

Combination of (51) and (56) produces

$$\lim_{n \to \infty} \frac{E \sum_{i=(1-\theta_s)n+1}^{n} \tilde{\mathbf{H}}_i}{n} = \frac{\sqrt{2}\Gamma(\frac{d+1}{2})}{\Gamma(\frac{d}{2})} \left(1 - \gamma_{inc}(\gamma_{inc}^{-1}(1-\theta_s, \frac{d}{2}), \frac{d+1}{2})\right). \tag{57}$$

In a completely analogous way we obtain

$$\lim_{n \to \infty} \frac{E \sum_{i=(1-\beta_s)n+1}^{n} \tilde{\mathbf{H}}_i}{n} = \frac{\sqrt{2}\Gamma(\frac{d+1}{2})}{\Gamma(\frac{d}{2})} \left(1 - \gamma_{inc}(\gamma_{inc}^{-1}(1-\beta_s, \frac{d}{2}), \frac{d+1}{2})\right). \tag{58}$$

Similarly to (54) we easily determine

$$\frac{2^{1-\frac{k}{2}}}{\Gamma(\frac{d}{2})}\int_0^{F_{\chi_d}^{-1}((1+\epsilon)\delta_s)} t^{d-1}e^{-\frac{t^2}{2}}dt = (1+\epsilon)\delta_s$$

$$\implies F_{\chi_d}^{-1}((1+\epsilon)\delta_s) = \sqrt{2\gamma_{inc}^{-1}((1+\epsilon)\delta_s,\frac{d}{2})} = \sqrt{2\gamma_{inc}^{-1}((1+\epsilon)(1-\theta_s),\frac{d}{2})} \quad (59)$$

Combination of (51), (57), (58), and (59) gives us the following equation for computing $\theta_s$

$$(1-\epsilon)\frac{\frac{\sqrt{2}\Gamma(\frac{d+1}{2})}{\Gamma(\frac{d}{2})}\left((1-\gamma_{inc}(\gamma_{inc}^{-1}(1-\theta_s,\frac{d}{2}),\frac{d+1}{2}))-2(1-\gamma_{inc}(\gamma_{inc}^{-1}(1-\beta_s,\frac{d}{2}),\frac{d+1}{2}))\right)}{\theta_s}-\sqrt{2\gamma_{inc}^{-1}((1+\epsilon)(1-\theta_s),\frac{d}{2})}=0. \quad (60)$$

Let $\hat{\theta}_s$ be the solution of (60). Then $\delta_s = 1 - \hat{\theta}_s$ and $c_s = \delta_s n = (1-\hat{\theta}_s)n$. This concludes step 1.

*Step* 2:

In this step we compute $\lim_{n\to\infty}\left(\frac{E\sum_{i=c_s+1}^n \tilde{\mathbf{H}}_i^2}{n} - \frac{(E(\tilde{\mathbf{H}}^T\mathbf{z})-E\sum_{i=1}^{c_s}\tilde{\mathbf{H}}_i)^2}{n(n-c_s)}\right)$ with $c_s = (1-\hat{\theta}_s)n$. Using the results from step 1 we easily find

$$\lim_{n\to\infty}\frac{(E(\tilde{\mathbf{H}}^T\mathbf{z})-E\sum_{i=1}^{c_s}\tilde{\mathbf{H}}_i)^2}{n(n-c_s)} = \frac{\left(\frac{\sqrt{2}\Gamma(\frac{d+1}{2})}{\Gamma(\frac{d}{2})}((1-\gamma_{inc}(\gamma_{inc}^{-1}(1-\hat{\theta}_s,\frac{d}{2}),\frac{d+1}{2}))-2(1-\gamma_{inc}(\gamma_{inc}^{-1}(1-\beta_s,\frac{d}{2}),\frac{d+1}{2})))\right)^2}{\hat{\theta}_s}. \quad (61)$$

Effectively, what is left to compute is $\lim_{n\to\infty}\frac{E\sum_{i=c_s+1}^n \tilde{\mathbf{H}}_i^2}{n}$. Using an approach similar to the one used in step 1 and following [8, 77] we have

$$\lim_{n\to\infty}\frac{E\sum_{i=(1-\hat{\theta}_s)n+1}^n \tilde{\mathbf{H}}_i^2}{n} = \int_{F_{\chi_d^2}^{-1}(1-\hat{\theta}_s)}^\infty t\,dF_{\chi_d^2}(t) \quad (62)$$

where $F_{\chi_d^2}(\cdot)$ is the cdf of the chi-square random variable with $d$ degrees of freedom and naturally $F_{\chi_d^2}^{-1}(\cdot)$ is the inverse cdf of the chi-square random variable with $d$ degrees of freedom. We then have

$$dF_{\chi_d^2}(t) = \frac{2^{-\frac{d}{2}}}{\Gamma(\frac{d}{2})}t^{\frac{d}{2}-1}e^{-\frac{t}{2}}, t\geq 0 \quad (63)$$

where as earlier $\Gamma(\cdot)$ stands for the gamma function. The following integration then gives us $F_{\chi_d^2}^{-1}(1-\hat{\theta}_s)$.

Namely,

$$\frac{2^{-\frac{d}{2}}}{\Gamma(\frac{d}{2})} \int_0^{F_{\chi_d^2}^{-1}(1-\hat{\theta}_s)} t^{\frac{d}{2}-1} e^{-\frac{t}{2}} dt = 1 - \hat{\theta}_s$$
$$\implies F_{\chi_d^2}^{-1}(1-\hat{\theta}_s) = 2\gamma_{inc}^{-1}(1-\hat{\theta}_s, \frac{d}{2}), \tag{64}$$

where as earlier $\gamma_{inc}^{-1}(\cdot, \cdot)$ is the inverse incomplete gamma function. We then find

$$\int_{F_{\chi_d^2}^{-1}(1-\hat{\theta}_s)}^{\infty} t dF_{\chi_d^2}(t) = \frac{2^{-\frac{k}{2}}}{\Gamma(\frac{d}{2})} \int_{F_{\chi_d^2}^{-1}(1-\hat{\theta}_s)}^{\infty} t^{\frac{d+2}{2}-1} e^{-\frac{t}{2}} dF_{\chi_d^2}(t) = \frac{2\Gamma(\frac{d+2}{2})}{\Gamma(\frac{d}{2})} \left(1 - \gamma_{inc}\left(\frac{F_{\chi_d^2}^{-1}(1-\hat{\theta}_s)}{2}, \frac{d+2}{2}\right)\right) \tag{65}$$

where as earlier $\gamma_{inc}(\cdot, \cdot)$ stands for the incomplete gamma function. From (64) and (65) we obtain

$$\int_{F_{\chi_d^2}^{-1}(1-\hat{\theta}_s)}^{\infty} t dF_{\chi_d^2}(t) = \frac{2\Gamma(\frac{d+2}{2})}{\Gamma(\frac{d}{2})} \left(1 - \gamma_{inc}(\gamma_{inc}^{-1}(1-\hat{\theta}_s, \frac{d}{2}), \frac{d+2}{2})\right). \tag{66}$$

Combination of (62) and (66) produces

$$\lim_{n \to \infty} \frac{E \sum_{i=(1-\hat{\theta}_s)n+1}^{n} \tilde{\mathbf{H}}_i^2}{n} = \frac{2\Gamma(\frac{d+2}{2})}{\Gamma(\frac{d}{2})} \left(1 - \gamma_{inc}(\gamma_{inc}^{-1}(1-\hat{\theta}_s, \frac{d}{2}), \frac{d+2}{2})\right). \tag{67}$$

We summarize the results from this section in the following theorem.

**Theorem 3.** *(Strong threshold) Let $A$ be a $dm \times dn$ measurement matrix in (1) with the null-space uniformly distributed in the Grassmanian. Let the unknown $\mathbf{x}$ in (1) be $k$-block-sparse with the length of its blocks $d$. Let $k, m, n$ be large and let $\alpha = \frac{m}{n}$ and $\beta_s = \frac{k}{n}$ be constants independent of $m$ and $n$. Let $\gamma_{inc}(\cdot, \cdot)$ be the incomplete gamma function and let $\gamma_{inc}^{-1}(\cdot, \cdot)$ be the inverse of the incomplete gamma function. Further, let $\epsilon > 0$ be an arbitrarily small constant and $\hat{\theta}_s$, $(\beta_s \leq \hat{\theta}_s \leq 1)$ be the solution of*

$$(1-\epsilon)\frac{\frac{\sqrt{2}\Gamma(\frac{d+1}{2})}{\Gamma(\frac{d}{2})}\left((1-\gamma_{inc}(\gamma_{inc}^{-1}(1-\theta_s, \frac{d}{2}), \frac{d+1}{2})) - 2(1-\gamma_{inc}(\gamma_{inc}^{-1}(1-\beta_s, \frac{d}{2}), \frac{d+1}{2}))\right)}{\theta_s} - \sqrt{2\gamma_{inc}^{-1}((1+\epsilon)(1-\theta_s), \frac{d}{2})} = 0. \tag{68}$$

*If $\alpha$ and $\beta_s$ further satisfy*

$$\alpha d > \frac{2\Gamma(\frac{d+2}{2})}{\Gamma(\frac{d}{2})}\left(1 - \gamma_{inc}(\gamma_{inc}^{-1}(1-\hat{\theta}_s, \frac{d}{2}), \frac{d+2}{2})\right)$$
$$- \frac{\left(\frac{\sqrt{2}\Gamma(\frac{d+1}{2})}{\Gamma(\frac{d}{2})}((1-\gamma_{inc}(\gamma_{inc}^{-1}(1-\hat{\theta}_s, \frac{d}{2}), \frac{d+1}{2})) - 2(1-\gamma_{inc}(\gamma_{inc}^{-1}(1-\beta_s, \frac{d}{2}), \frac{d+1}{2})))\right)^2}{\hat{\theta}_s} \quad (69)$$

*then the solutions of (1) and (3) coincide with overwhelming probability.*

*Proof.* Follows from the previous discussion combining (6), (8), (31), (49), (50), (60), (61), and (67). □

The results for the strong threshold obtained from the above theorem for different block-lengths $d$ are presented on Figure 3. The case of large $d$ was considered in [78, 81] and is given for comparison as $d \to \infty$ on Figure 3 as well. (In Section 5 we will show how the results given in [78, 81] follow from the above presented analysis.) Increasing the block-length introduces so to say more structure on the unknown signals. One would then expect that recoverable thresholds should be higher as $d$ increases. Figure 3 hints that $\ell_2/\ell_1$-optimization algorithm from (3) possibly indeed recovers higher block-sparsity as the block length increases.

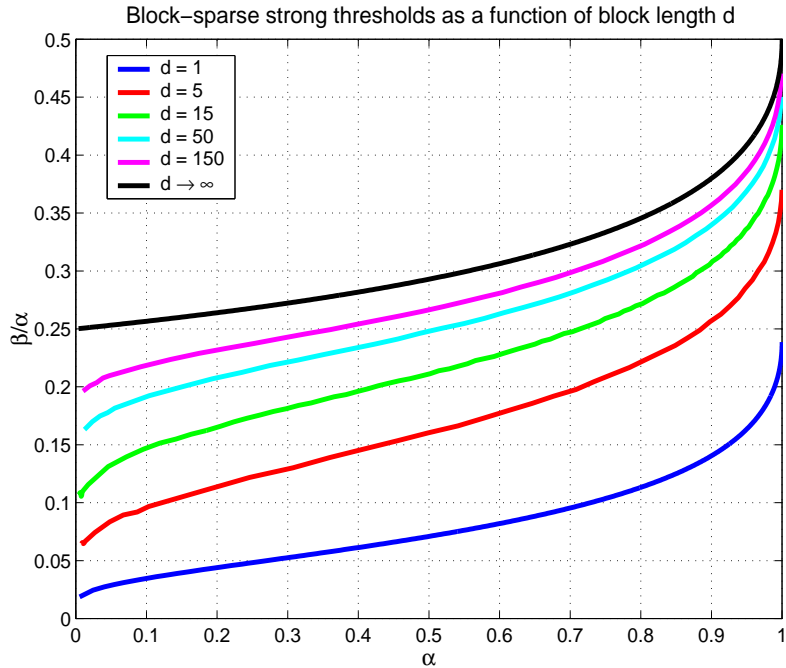

Figure 3: Block-sparse *strong* thresholds as a function of block-length $d$; $\ell_2/\ell_1$-optimization

## 4.2 Sectional threshold

In this subsection we determine the sectional threshold $\beta_{sec}$. Before proceeding further we quickly recall on the definition of the sectional threshold. Namely, for a given $\alpha$, $\beta_{sec}$ is the maximum value of $\beta$ such that the solutions of (1) and (3) coincide for any given $\beta n$-block-sparse $\mathbf{x}$ with a fixed location of nonzero blocks. Since the analysis that will follow will clearly be irrelevant with respect to what particular location of nonzero blocks are chosen, we can for the simplicity of the exposition and without loss of generality assume that the blocks $\mathbf{X}_1, \mathbf{X}_2, \ldots, \mathbf{X}_{n-k}$ of $\mathbf{x}$ are equal to zero (i.e. they are zero blocks). Under this assumption we have the following corollary of Theorem 1.

**Corollary 1** (Nonzero part of $\mathbf{x}$ has a fixed location). *Assume that a $dm \times dn$ measurement matrix $A$ is given. Let $\mathbf{x}$ be a $k$-block-sparse vector. Also let $\mathbf{X}_1 = \mathbf{X}_2 = \cdots = \mathbf{X}_{n-k} = 0$. Further, assume that $\mathbf{y} = A\mathbf{x}$ and that $\mathbf{w}$ is a $dn \times 1$ vector. Then (3) will produce the solution of (1) if*

$$(\forall \mathbf{w} \in \mathbf{R}^{dn} | A\mathbf{w} = 0) \quad \sum_{i=n-k+1}^{n} \|\mathbf{W}_i\|_2 < \sum_{i=1}^{n-k} \|\mathbf{W}_i\|_2. \tag{70}$$

Following the procedure of Subsection 4.1 we set $S_{sec}$

$$S_{sec} = \{\mathbf{w} \in S^{dn-1} | \sum_{i=n-k+1}^{n} \|\mathbf{W}_i\|_2 < \sum_{i=1}^{n-k} \|\mathbf{W}_i\|_2\} \tag{71}$$

and

$$w(S_{sec}) = E \sup_{\mathbf{w} \in S_{sec}} (\mathbf{h}^T \mathbf{w}) \tag{72}$$

where as earlier $\mathbf{h}$ is a random column vector in $R^{dn}$ with i.i.d. $\mathcal{N}(0,1)$ components and $S^{dn-1}$ is the unit $dn$-dimensional sphere. As in Subsection 4.1 our goal will be to compute an upper bound on $w(S_{sec})$ and then equal that upper bound to $\left(\sqrt{dm} - \frac{1}{4\sqrt{dm}}\right)$. In the following subsections we present a way to get such an upper bound. As earlier, we set $w(\mathbf{h}, S_{sec}) = \max_{\mathbf{w} \in S_{sec}} (\mathbf{h}^T \mathbf{w})$. Following the strategy of the previous sections in Subsection 4.2.1 we determine an upper bound $B_{sec}$ on $w(\mathbf{h}, S_{sec})$. In Subsection 4.2.2 we will compute an upper bound on $E(B_{sec})$. That quantity will be an upper bound on $w(S_{sec})$ since according to the following $E(B_{sec})$ is an upper bound on $w(S_{sec})$

$$w(S_{sec}) = Ew(\mathbf{h}, S_{sec}) = E(\max_{\mathbf{w} \in S_{sec}} (\mathbf{h}^T \mathbf{w})) \leq E(B_{sec}). \tag{73}$$

### 4.2.1 Upper-bounding $w(\mathbf{h}, S_{sec})$

The following sequence of equalities is analogous to (9)

$$w(\mathbf{h}, S_{sec}) = \max_{\mathbf{w}\in S_{sec}} (\mathbf{h}^T\mathbf{w}) = \max_{\mathbf{w}\in S_{sec}} \sum_{i=1}^{n} |\mathbf{h}_i\mathbf{w}_i| = \max_{\mathbf{w}\in S_{sec}} \sum_{i=1}^{n} \|\mathbf{H}_i\|_2\|\mathbf{W}_i\|_2. \quad (74)$$

Let $\mathbf{H}_{(norm)}^{(n-k)} = (\|\mathbf{H}_1\|_2, \|\mathbf{H}_2\|_2, \ldots, \|\mathbf{H}_{n-k}\|_2)$. Further, let $|\mathbf{H}_{(norm)}^{(n-k)}|_{(i)}$ be the $i$-th smallest of the elements of $\mathbf{H}_{(norm)}^{(n-k)}$. Set

$$\hat{\mathbf{H}} = (|\mathbf{H}_{(norm)}^{(n-k)}|_{(1)}, |\mathbf{H}_{(norm)}^{(n-k)}|_{(2)}, \ldots, |\mathbf{H}_{(norm)}^{(n-k)}|_{(n-k)}, \|\mathbf{H}_{n-k+1}\|_2, \|\mathbf{H}_{n-k+2}\|_2, \ldots, \|\mathbf{H}_n\|_2)^T. \quad (75)$$

If $\mathbf{w} \in S_{sec}$ then a vector obtained by permuting the blocks of $\mathbf{w}$ in any possible way is also in $S_{sec}$. Then (74) can be rewritten as

$$w(\mathbf{h}, S_{sec}) = \max_{\mathbf{w}\in S_{sec}} \sum_{i=1}^{n} \hat{\mathbf{H}}_i \|\mathbf{W}_i\|_2 \quad (76)$$

where $\hat{\mathbf{H}}_i$ is the $i$-th element of vector $\hat{\mathbf{H}}$. Let $\mathbf{y} = (\mathbf{y}_1, \mathbf{y}_2, \ldots, \mathbf{y}_n)^T \in R^n$. Then one can simplify (76) in the following way

$$\begin{aligned}
w(\mathbf{h}, S_{sec}) = \max_{\mathbf{y}\in R^n} \quad & \sum_{i=1}^{n} \hat{\mathbf{H}}_i \mathbf{y}_i \\
\text{subject to} \quad & \mathbf{y}_i \geq 0, 0 \leq i \leq n \\
& \sum_{i=n-k+1}^{n} \mathbf{y}_i \geq \sum_{i=1}^{n-k} \mathbf{y}_i \\
& \sum_{i=1}^{n} \mathbf{y}_i^2 \leq 1.
\end{aligned} \quad (77)$$

One can then proceed in a fashion similar to the one from Subsection 4.1.1 and compute an upper bound based on duality. The only difference is that we now have $\hat{\mathbf{H}}$ instead of $\tilde{\mathbf{H}}$. After repeating literally every step of the derivation from Subsection 4.1.1 one obtains the following analogue to the equation (30)

$$w(\mathbf{h}, S_{sec}) \leq \sqrt{\sum_{i=1}^{n} \hat{\mathbf{H}}_i^2 - \sum_{i=1}^{c} \hat{\mathbf{H}}_i^2 - \frac{((\hat{\mathbf{H}}^T\mathbf{z}) - \sum_{i=1}^{c} \hat{\mathbf{H}}_i)^2}{n-c}} = \sqrt{\sum_{i=c+1}^{n} \hat{\mathbf{H}}_i^2 - \frac{((\hat{\mathbf{H}}^T\mathbf{z}) - \sum_{i=1}^{c} \hat{\mathbf{H}}_i)^2}{n-c}} \quad (78)$$

where $c \leq (n-k)$ is such that $((\hat{\mathbf{H}}^T\mathbf{z}) - \sum_{i=1}^{c} \hat{\mathbf{H}}_i) \geq 0$. As earlier, as long as $(\hat{\mathbf{H}}^T\mathbf{z}) \geq 0$ there will be a $c$ (it is possible that $c=0$) such that quantity on the most right hand side of (78) is an upper bound on $w(\mathbf{h}, S_{sec})$.

Using (78) we then establish the following analogue to Lemma 1.

**Lemma 4.** *Let $\mathbf{h} \in R^n$ be a vector with i.i.d. zero-mean unit variance gaussian components. Further let $\hat{\mathbf{H}}$ be as defined in (75) and $w(\mathbf{h}, S_{sec}) = \max_{\mathbf{w} \in S_{sec}}(\mathbf{h}^T \mathbf{w})$ where $S_{sec}$ is as defined in (71). Let $\mathbf{z} \in R^n$ be a column vector such that $\mathbf{z}_i = 1, 1 \leq i \leq (n-k)$ and $\mathbf{z}_i = -1, n-k+1 \leq i \leq n$. Then*

$$w(\mathbf{h}, S_{sec}) \leq B_{sec} \tag{79}$$

*where*

$$B_{sec} = \begin{cases} \sqrt{\sum_{i=1}^{n} \hat{\mathbf{H}}_i^2} & \text{if } \zeta_{sec}(\mathbf{h}, c_{sec}) \leq 0 \\ \sqrt{\sum_{i=c_{sec}+1}^{n} \hat{\mathbf{H}}_i^2 - \frac{((\hat{\mathbf{H}}^T \mathbf{z}) - \sum_{i=1}^{c_{sec}} \hat{\mathbf{H}}_i)^2}{n - c_{sec}}} & \text{if } \zeta_{sec}(\mathbf{h}, c_{sec}) > 0 \end{cases}, \tag{80}$$

$\zeta_{sec}(\mathbf{h}, c) = \frac{(\hat{\mathbf{H}}^T \mathbf{z}) - \sum_{i=1}^{c} \hat{\mathbf{H}}_i}{n-c} - \hat{\mathbf{H}}_c$ *and* $c_{sec} = \delta_{sec} n$ *is a $c \leq n - k$ such that*

$$\frac{(1-\epsilon)E((\hat{\mathbf{H}}^T \mathbf{z}) - \sum_{i=1}^{c} \hat{\mathbf{H}}_i)}{n-c} - F_{\chi_d}^{-1}\left(\frac{(1+\epsilon)c}{n-k}\right) = 0. \tag{81}$$

$F_{\chi_d}^{-1}(\cdot)$ *is the inverse cdf of the chi random variable with $d$ degrees of freedom. $\epsilon > 0$ is an arbitrarily small constant independent of $n$.*

*Proof.* Follows directly from the derivation before Lemma 1. □

### 4.2.2 Computing an upper bound on $E(B_{sec})$

Following step-by-step the derivation of Lemma 3 (with a trivial adjustment in finding Lipschitz constant $\sigma$) we can establish the sectional threshold analogue to it.

**Lemma 5.** *Assume the setup of Lemma 4. Let further $\psi_{sec} = \frac{E(\hat{\mathbf{H}}^T \mathbf{z}) - \sum_{i=1}^{c_{sec}} \hat{\mathbf{H}}_i}{n}$. Then*

$$E(B_{sec}) \leq \sqrt{n}\left(\exp\left\{-\frac{n\epsilon^2 \delta_{sec}}{2(1+\epsilon)}\right\} + \exp\left\{-\frac{\epsilon^2 \psi_{sec}^2 n}{2}\right\}\right) + \sqrt{E \sum_{i=c_{sec}+1}^{n} \hat{\mathbf{H}}_i^2 - \frac{(E(\hat{\mathbf{H}}^T \mathbf{z}) - E \sum_{i=1}^{c_{sec}} \hat{\mathbf{H}}_i)^2}{n - c_{sec}}}. \tag{82}$$

*Proof.* Follows directly from the derivation before Lemma 3. □

Similarly to (50), if $n$ is large, for a fixed $\alpha$ one can determine $\beta_{sec}$ as a maximum $\beta$ such that

$$\alpha d > \frac{E \sum_{i=c_{sec}+1}^{n} \hat{\mathbf{H}}_i^2}{n} - \frac{(E(\hat{\mathbf{H}}^T \mathbf{z}) - E \sum_{i=1}^{c_{sec}} \hat{\mathbf{H}}_i)^2}{n(n - c_{sec})}. \tag{83}$$

In the rest of this subsection we show how the left hand side of (83) can be computed for a randomly chosen fixed $\beta_{sec}$. We again, as earlier, do so in two steps:

1. We first determine $c_{sec}$

2. We then compute $\lim_{n\to\infty}\left(\frac{E\sum_{i=c_{sec}+1}^{n}\hat{\mathbf{H}}_i^2}{n} - \frac{(E(\hat{\mathbf{H}}^T\mathbf{z}) - E\sum_{i=1}^{c_{sec}}\hat{\mathbf{H}}_i)^2}{n(n-c_{sec})}\right)$ with $c_{sec}$ found in step 1.

*Step 1:*

From Lemma 4 we have $c_{sec} = \delta_{sec} n$ is a $c$ such that

$$\frac{(1-\epsilon)E((\sum_{i=1}^{n-\beta_{sec}n}\hat{\mathbf{H}}_i - \sum_{i=n-\beta_{sec}n+1}^{n}\hat{\mathbf{H}}_i) - \sum_{i=1}^{\delta_{sec}n}\hat{\mathbf{H}}_i)}{n-c} - F_{\chi_d}^{-1}\left(\frac{(1+\epsilon)c}{n(1-\beta_{sec})}\right) = 0$$

$$\Leftrightarrow \frac{(1-\epsilon)(E\sum_{i=1}^{n-\beta_{sec}n}\hat{\mathbf{H}}_i - E\sum_{i=n-\beta_{sec}n+1}^{n}\|\mathbf{H}_i\|_2 - E\sum_{i=1}^{\delta_{sec}n}\hat{\mathbf{H}}_i)}{n-c} - F_{\chi_d}^{-1}\left(\frac{(1+\epsilon)c}{n(1-\beta_{sec})}\right) = 0 \quad (84)$$

where as earlier $\hat{\mathbf{H}}_i = |\mathbf{H}_{(norm)}^{(n-k)}|_{(i)}, 1 \leq i \leq (n-\beta_{sec}n)$, is the $i$-th smallest magnitude of blocks $\mathbf{H}_i, 1 \leq 1 \leq 1 : (n-\beta_{sec}n)$. We also recall that $\|\mathbf{H}_i\|_2, n-\beta_{sec}n+1 \leq i \leq n$, are the magnitudes of the last $\beta_{sec}n$ blocks of vector $\mathbf{h}$ (these magnitudes of last $\beta_{sec}n$ blocks of vector $\mathbf{h}$ are not sorted). As earlier, all components of $\mathbf{h}$ are i.i.d. zero-mean unit variance Gaussian random variables and $\epsilon > 0$ is an arbitrarily small constant. Then since $\|\mathbf{H}_i\|_2$ is a chi-distributed random variable with $d$ degrees of freedom we clearly have $E\|\mathbf{H}_i\|_2 = \frac{\sqrt{2}\Gamma(\frac{d+1}{2})}{\Gamma(\frac{d}{2})}, n-\beta_{sec}n+1 \leq i \leq n$. Then from (84)

$$\frac{(1-\epsilon)E((\sum_{i=1}^{n-\beta_{sec}n}\hat{\mathbf{H}}_i - \sum_{i=n-\beta_{sec}n+1}^{n}\hat{\mathbf{H}}_i) - \sum_{i=1}^{\delta_{sec}n}\hat{\mathbf{H}}_i)}{n-c} - F_{\chi_d}^{-1}\left(\frac{(1+\epsilon)c}{n(1-\beta_{sec})}\right) = 0$$

$$\Leftrightarrow \frac{(1-\epsilon)(E\sum_{i=\delta_{sec}n+1}^{n-\beta_{sec}n}\hat{\mathbf{H}}_i - \frac{\sqrt{2}\Gamma(\frac{d+1}{2})}{\Gamma(\frac{d}{2})}\beta_{sec}n)}{n(1-\delta_{sec})} - F_{\chi_d}^{-1}\left(\frac{(1+\epsilon)\delta_{sec}n}{n(1-\beta_{sec})}\right) = 0. \quad (85)$$

Set $\theta_{sec} = 1 - \delta_{sec}$. Following the derivation of (57) we have

$$\lim_{n\to\infty} \frac{E\sum_{i=(1-\theta_{sec})n+1}^{(1-\beta_{sec})n}\tilde{\mathbf{H}}_i}{n(1-\beta_{sec})} = \frac{\sqrt{2}\Gamma(\frac{d+1}{2})}{\Gamma(\frac{d}{2})}\left(1 - \gamma_{inc}(\gamma_{inc}^{-1}(\frac{1-\theta_{sec}}{1-\beta_{sec}},\frac{d}{2}),\frac{d+1}{2})\right). \quad (86)$$

Similarly to (59) we easily determine

$$F_{\chi_d}^{-1}\left(\frac{(1+\epsilon)(1-\theta_{sec})}{1-\beta_{sec}}\right) = \sqrt{2\gamma_{inc}^{-1}(\frac{(1+\epsilon)(1-\theta_{sec})}{1-\beta_{sec}},\frac{d}{2})} \quad (87)$$

Combination of (84), (85), (86), and (87) gives us the following equation for computing $\theta_{sec}$

$$(1-\epsilon)\frac{(1-\beta_{sec})\frac{\sqrt{2}\Gamma(\frac{d+1}{2})}{\Gamma(\frac{d}{2})}\left(1-\gamma_{inc}(\gamma_{inc}^{-1}(\frac{1-\theta_{sec}}{1-\beta_{sec}},\frac{d}{2}),\frac{d+1}{2})\right)-\frac{\sqrt{2}\Gamma(\frac{d+1}{2})}{\Gamma(\frac{d}{2})}\beta_{sec}}{\theta_{sec}}-\sqrt{2\gamma_{inc}^{-1}(\frac{(1+\epsilon)(1-\theta_{sec})}{1-\beta_{sec}},\frac{d}{2})}=0. \tag{88}$$

Let $\hat{\theta}_{sec}$ be the solution of (88). Then $\delta_{sec} = 1 - \hat{\theta}_{sec}$ and $c_{sec} = \delta_{sec}n = (1-\hat{\theta}_{sec})n$. This concludes step 1.

*Step 2:*

In this step we compute $\lim_{n\to\infty}\left(\frac{E\sum_{i=c_{sec}+1}^{n}\hat{\mathbf{H}}_i^2}{n}-\frac{(E(\hat{\mathbf{H}}^T\mathbf{z})-E\sum_{i=1}^{c_{sec}}\hat{\mathbf{H}}_i)^2}{n(n-c_{sec})}\right)$ with $c_{sec} = (1-\hat{\theta}_{sec})n$. Using results from step 1 we easily find

$$\lim_{n\to\infty}\frac{(E(\hat{\mathbf{H}}^T\mathbf{z})-E\sum_{i=1}^{c_{sec}}\hat{\mathbf{H}}_i)^2}{n(n-c_{sec})}=\frac{\left((1-\beta_{sec})\frac{\sqrt{2}\Gamma(\frac{d+1}{2})}{\Gamma(\frac{d}{2})}(1-\gamma_{inc}(\gamma_{inc}^{-1}(\frac{1-\hat{\theta}_{sec}}{1-\beta_{sec}},\frac{d}{2}),\frac{d+1}{2}))-\frac{\sqrt{2}\Gamma(\frac{d+1}{2})}{\Gamma(\frac{d}{2})}\beta_{sec}\right)^2}{\hat{\theta}_{sec}}. \tag{89}$$

What is left to compute is $\lim_{n\to\infty}\frac{E\sum_{i=c_{sec}+1}^{n}\hat{\mathbf{H}}_i^2}{n}$. We first observe

$$\frac{E\sum_{i=c_{sec}+1}^{n}\hat{\mathbf{H}}_i^2}{n}=\frac{E\sum_{i=c_{sec}+1}^{(1-\beta_{sec})n}\hat{\mathbf{H}}_i^2}{n}+\frac{E\sum_{i=(1-\beta_{sec})n+1}^{n}\hat{\mathbf{H}}_i^2}{n}=\frac{E\sum_{i=(1-\hat{\theta}_{sec})n+1}^{(1-\beta_{sec})n}\hat{\mathbf{H}}_i^2}{n}+\beta_{sec}d. \tag{90}$$

Following the derivation of (67) we also have

$$\lim_{n\to\infty}\frac{E\sum_{i=(1-\hat{\theta}_{sec})n+1}^{(1-\beta_{sec})n}\hat{\mathbf{H}}_i^2}{n(1-\beta_{sec})}=\frac{2\Gamma(\frac{d+2}{2})}{\Gamma(\frac{d}{2})}\left(1-\gamma_{inc}(\gamma_{inc}^{-1}(\frac{1-\hat{\theta}_{sec}}{1-\beta_{sec}},\frac{d}{2}),\frac{d+2}{2})\right). \tag{91}$$

Combining (90) and (91) we find

$$\lim_{n\to\infty}\frac{E\sum_{i=c_{sec}+1}^{n}\hat{\mathbf{H}}_i^2}{n}=(1-\beta_{sec})\frac{2\Gamma(\frac{d+2}{2})}{\Gamma(\frac{d}{2})}\left(1-\gamma_{inc}(\gamma_{inc}^{-1}(\frac{1-\hat{\theta}_{sec}}{1-\beta_{sec}},\frac{d}{2}),\frac{d+2}{2})\right)+\beta_{sec}d. \tag{92}$$

We summarize the results from this section in the following theorem.

**Theorem 4.** *(Sectional threshold) Let $A$ be a $dm \times dn$ measurement matrix in (1) with the null-space uniformly distributed in the Grassmanian. Let the unknown $\mathbf{x}$ in (1) be $k$-block-sparse with the length of its blocks $d$. Further, let the location of nonzero blocks of $\mathbf{x}$ be arbitrarily chosen but fixed. Let $k, m, n$ be large and let $\alpha = \frac{m}{n}$ and $\beta_{sec} = \frac{k}{n}$ be constants independent of $m$ and $n$. Let $\gamma_{inc}(\cdot,\cdot)$ and $\gamma_{inc}^{-1}(\cdot,\cdot)$ be the incomplete gamma function and its inverse, respectively. Further, let $\epsilon > 0$ be an arbitrarily small constant* and $\hat{\theta}_{sec}$, ($\beta_{sec} \leq \hat{\theta}_{sec} \leq 1$) be the solution of

$$(1-\epsilon)\frac{(1-\beta_{sec})\frac{\sqrt{2}\Gamma(\frac{d+1}{2})}{\Gamma(\frac{d}{2})}\left(1-\gamma_{inc}(\gamma_{inc}^{-1}(\frac{1-\theta_{sec}}{1-\beta_{sec}},\frac{d}{2}),\frac{d+1}{2})\right) - \frac{\sqrt{2}\Gamma(\frac{d+1}{2})}{\Gamma(\frac{d}{2})}\beta_{sec}}{\theta_{sec}} - \sqrt{2\gamma_{inc}^{-1}(\frac{(1+\epsilon)(1-\theta_{sec})}{1-\beta_{sec}},\frac{d}{2})} = 0. \quad (93)$$

*If $\alpha$ and $\beta_{sec}$ further satisfy*

$$\alpha d > (1-\beta_{sec})\frac{2\Gamma(\frac{d+2}{2})}{\Gamma(\frac{d}{2})}\left(1-\gamma_{inc}(\gamma_{inc}^{-1}(\frac{1-\hat{\theta}_{sec}}{1-\beta_{sec}},\frac{d}{2}),\frac{d+2}{2})\right) + \beta_{sec}d$$

$$-\frac{\left((1-\beta_{sec})\frac{\sqrt{2}\Gamma(\frac{d+1}{2})}{\Gamma(\frac{d}{2})}(1-\gamma_{inc}(\gamma_{inc}^{-1}(\frac{1-\hat{\theta}_{sec}}{1-\beta_{sec}},\frac{d}{2}),\frac{d+1}{2})) - \frac{\sqrt{2}\Gamma(\frac{d+1}{2})}{\Gamma(\frac{d}{2})}\beta_{sec}\right)^2}{\hat{\theta}_{sec}} \quad (94)$$

*then the solutions of (1) and (3) coincide with overwhelming probability.*

*Proof.* Follows from the previous discussion combining (6), (73), (79), (82), (83), (88), (89), and (92). □

The results for the sectional threshold obtained from the above theorem for different block-lengths $d$ are presented on Figure 4. We also show on Figure 4 the results from [78, 81] when $d \to \infty$. (These results were derived for the strong threshold; however, any lower bound on the strong threshold is automatically a lower bound on the sectional threshold as well.) In the following section we will explicitly show how the results shown on Figure 4 for $d \to \infty$ follow from the derivation given above.

### 4.3 Weak threshold

In this subsection we determine the weak threshold $\beta_w$. Before proceeding further we again quickly recall on the definition of the weak threshold. Namely, for a given $\alpha$, $\beta_w$ is the maximum value of $\beta$ such that the solutions of (1) and (3) coincide for any $\beta n$-block-sparse $\mathbf{x}$ with a given fixed location of non-zero blocks and given fixed directions of non-zero block vectors $\mathbf{X}_i$. As in Subsection 4.2 we can for the simplicity of the exposition and without loss of generality assume that the blocks $\mathbf{X}_1, \mathbf{X}_2, \ldots, \mathbf{X}_{n-k}$ of $\mathbf{x}$ are equal to zero and that that vectors $\mathbf{X}_{n-k+1}, \mathbf{X}_{n-k+2}, \ldots, \mathbf{X}_n$ have fixed directions. Furthermore, since all probability distributions of interest will be rotationally invariant we will later assume that $\mathbf{X}_i = (\|\mathbf{X}_i\|_2, 0, 0, \ldots, 0), n-k+1 \leq i \leq n$. We first have the following corollary of Theorem 1.

**Corollary 2.** *(Nonzero blocks of $\mathbf{x}$ have fixed directions and location) Assume that a $dm \times dn$ measurement matrix $A$ is given. Let $\mathbf{x}$ be a $k$-block-sparse vector. Also let $\mathbf{X}_1 = \mathbf{X}_2 = \cdots = \mathbf{X}_{n-k} = 0$. Let the directions of vectors $\mathbf{X}_{n-k+1}, \mathbf{X}_{n-k+2}, \ldots, \mathbf{X}_n$ be fixed. Further, assume that $\mathbf{y} = A\mathbf{x}$ and that $\mathbf{w}$ is a*

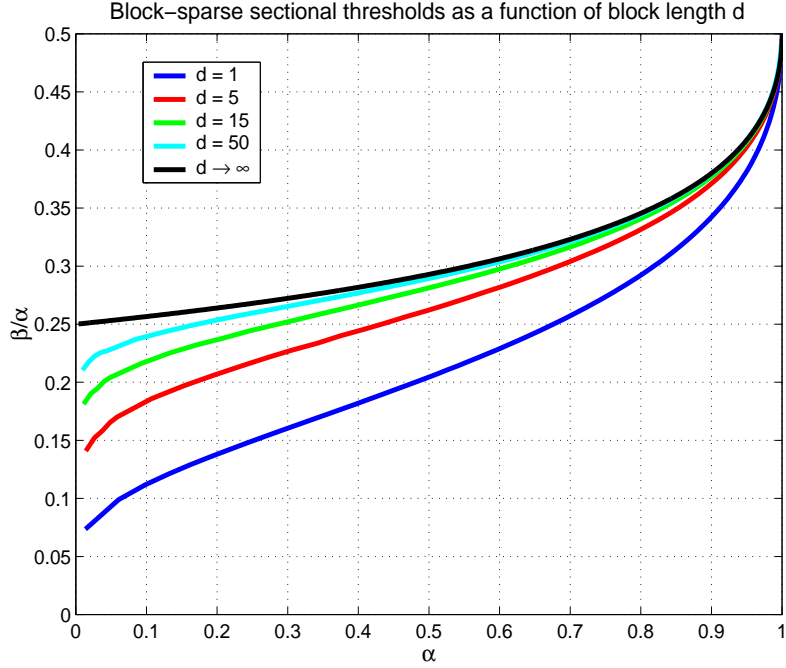

Figure 4: Block-sparse *sectional* thresholds as a function of block length $d$, $\ell_2/\ell_1$-optimization

$dn \times 1$ *vector. Then (3) will produce the solution of (1) if*

$$(\forall \mathbf{w} \in \mathbf{R}^{dn} | A\mathbf{w} = 0) \quad - \sum_{i=n-k+1}^{n} \frac{\mathbf{X}_i^T \mathbf{W}_i}{\|\mathbf{X}_i\|_2} < \sum_{i=1}^{n-k} \|\mathbf{W}_i\|_2. \tag{95}$$

*Proof.* The proof closely follows the proof of Theorem 1 given in [83]. Let $\bar{\mathbf{x}}$ be the solution of (1) and let $\hat{\mathbf{x}}$ be the solution of (3). Also, assume $\bar{\mathbf{x}} \neq \hat{\mathbf{x}}$, i.e. assume $\sum_{i=1}^{n} \|\hat{\mathbf{X}}_i\|_2 \leq \sum_{i=1}^{n} \|\bar{\mathbf{X}}_i\|_2$ where $\bar{\mathbf{X}}_i = (\bar{\mathbf{x}}_{(i-1)d+1}, \bar{\mathbf{x}}_{(i-1)d+2}, \ldots, \bar{\mathbf{x}}_{id})^T$ and $\hat{\mathbf{X}}_i = (\hat{\mathbf{x}}_{(i-1)d+1}, \hat{\mathbf{x}}_{(i-1)d+2}, \ldots, \hat{\mathbf{x}}_{id})^T$, for $i = 1, 2, \ldots, n$. Then we can write

$$\begin{aligned}
\sum_{i=1}^{n} \|\hat{\mathbf{X}}_i\|_2 &= \sum_{i=1}^{n} \|\hat{\mathbf{X}}_i - \bar{\mathbf{X}}_i + \bar{\mathbf{X}}_i\|_2 = \sum_{i=n-k+1}^{n} \|\hat{\mathbf{X}}_i - \bar{\mathbf{X}}_i + \bar{\mathbf{X}}_i\|_2 + \sum_{i=1}^{n-k} \|\hat{\mathbf{X}}_i - \bar{\mathbf{X}}_i + \bar{\mathbf{X}}_i\|_2 \\
&= \sum_{i=n-k+1}^{n} \|\mathbf{W}_i + \bar{\mathbf{X}}_i\|_2 + \sum_{i=1}^{n-k} \|\mathbf{W}_i\|_2 \geq \sum_{i=n-k+1}^{n} \|\|\bar{\mathbf{X}}_i\|_2 + \frac{\bar{\mathbf{X}}_i^T \mathbf{W}_i}{\|\bar{\mathbf{X}}_i\|_2 \|\bar{\mathbf{W}}_i\|_2} \|\bar{\mathbf{W}}_i\|_2 | + \sum_{i=1}^{n-k} \|\mathbf{W}_i\|_2 \\
&\geq \sum_{i=n-k+1}^{n} \|\bar{\mathbf{X}}_i\|_2 + \sum_{i=n-k+1}^{n} \frac{\bar{\mathbf{X}}_i^T \mathbf{W}_i}{\|\bar{\mathbf{X}}_i\|_2} + \sum_{i=1}^{n-k} \|\mathbf{W}_i\|_2 = \sum_{i=1}^{n} \|\bar{\mathbf{X}}_i\|_2 + \sum_{i=n-k+1}^{n} \frac{\bar{\mathbf{X}}_i^T \mathbf{W}_i}{\|\bar{\mathbf{X}}_i\|_2} + \sum_{i=1}^{n-k} \|\mathbf{W}_i\|_2.
\end{aligned} \tag{96}$$

If (95) holds then from (96) $\sum_{i=1}^{n} ||\hat{\mathbf{X}}_i||_2 > \sum_{i=1}^{n} ||\bar{\mathbf{X}}_i||_2$ which contradicts the assumption $\sum_{i=1}^{n} ||\hat{\mathbf{X}}_i||_2 \leq \sum_{i=1}^{n} ||\bar{\mathbf{X}}_i||_2$. Therefore, $\bar{\mathbf{x}} = \hat{\mathbf{x}}$. This concludes the proof. $\square$

Following the procedure of Subsection 4.2 we set

$$S'_w = \{\mathbf{w} \in S^{dn-1} | \quad -\sum_{i=n-k+1}^{n} \frac{\mathbf{X}_i^T \mathbf{W}_i}{||\mathbf{X}_i||_2} < \sum_{i=1}^{n-k} ||\mathbf{W}_i||_2\} \tag{97}$$

and

$$w(S'_w) = E \sup_{\mathbf{w} \in S'_w} (\mathbf{h}^T \mathbf{w}) \tag{98}$$

where as earlier $\mathbf{h}$ is a random column vector in $R^{dn}$ with i.i.d. $\mathcal{N}(0,1)$ components and $S^{dn-1}$ is the unit $dn$-dimensional sphere. Let $\Theta_i$ be the orthogonal matrices such that $\mathbf{X}_i^T \Theta_i = (||\mathbf{X}_i||_2, 0, \ldots, 0), n-k+1 \leq i \leq n$. Set

$$S_w = \{\mathbf{w} \in S^{dn-1} | \quad -\sum_{i=n-k+1}^{n} \mathbf{w}_{(i-1)d+1} < \sum_{i=1}^{n-k} ||\mathbf{W}_i||_2\} \tag{99}$$

and

$$w(S_w) = E \sup_{\mathbf{w} \in S_w} (\mathbf{h}^T \mathbf{w}). \tag{100}$$

Since $\mathbf{H}_i^T$ and $\mathbf{H}_i^T \Theta_i$ have the same distribution we have $w(S_w) = w(S'_w)$. As in Subsections 4.1 and 4.2 our goal will then again be to compute an upper bound on $w(S_w)$ and subsequently equal that upper bound to $\left(\sqrt{dm} - \frac{1}{4\sqrt{dm}}\right)$. Following the strategy of the previous sections in Subsection 4.3.1 we will determine an upper bound $B_w$ on $w(\mathbf{h}, S_w)$. In Subsection 4.3.2 we will compute an upper bound on $E(B_w)$. That quantity will be an upper bound on $w(S_w)$ since according to the following $E(B_w)$ is an upper bound on $w(S_w)$

$$w(S_w) = Ew(\mathbf{h}, S_w) = E(\max_{\mathbf{w} \in S_w} (\mathbf{h}^T \mathbf{w})) \leq E(B_w). \tag{101}$$

### 4.3.1 Upper-bounding $w(\mathbf{h}, S_w)$

Let $\mathbf{H}_i^* = (\mathbf{h}_{(i-1)d+2}, \mathbf{h}_{(i-1)d+3}, \ldots, \mathbf{h}_{id})^T$, $\mathbf{W}_i^* = (\mathbf{w}_{(i-1)d+2}, \mathbf{w}_{(i-1)d+3}, \ldots, \mathbf{w}_{id})^T$, $i = n - k + 1, 2, \ldots, n$. One then writes in a way analogous to (9)

$$w(\mathbf{h}, S_w) = \max_{\mathbf{w} \in S_w} (\mathbf{h}^T \mathbf{w}) = \max_{\mathbf{w} \in S_w} (\sum_{i=n-k+1}^{n} \mathbf{h}_{(i-1)d+1} \mathbf{w}_{(i-1)d+1} + \sum_{i=n-k+1}^{n} ||\mathbf{H}_i^*||_2 ||\mathbf{W}_i^*||_2 + \sum_{i=1}^{n-k} ||\mathbf{H}_i||_2 ||\mathbf{W}_i||_2). \tag{102}$$

We recall one more time that $\mathbf{H}_{(norm)}^{(n-k)} = (\|\mathbf{H}_1\|_2, \|\mathbf{H}_2\|_2, \ldots, \|\mathbf{H}_{n-k}\|_2)$ and that $|\mathbf{H}_{(norm)}^{(n-k)}|_{(i)}$ is the $i$-th smallest of the elements of $\mathbf{H}_{(norm)}^{(n-k)}$. Set

$$\bar{\mathbf{H}} = (|\mathbf{H}_{(norm)}^{(n-k)}|_{(1)}, |\mathbf{H}_{(norm)}^{(n-k)}|_{(2)}, \ldots, |\mathbf{H}_{(norm)}^{(n-k)}|_{(n-k)}, -\mathbf{h}_{(n-k+1)d+1}, -\mathbf{h}_{(n-k+2)d+1}, \ldots, -\mathbf{h}_{(n-1)d+1},$$
$$\|\mathbf{H}_{n-k+1}^*\|_2, \|\mathbf{H}_{n-k+2}^*\|_2, \ldots, \|\mathbf{H}_n^*\|_2)^T. \quad (103)$$

Let $\bar{\mathbf{y}} = (\mathbf{y}_1, \mathbf{y}_2, \ldots, \mathbf{y}_{n+k})^T \in R^{n+k}$. Then one can simplify (102) in the following way

$$w(\mathbf{h}, S_w) = \max_{\bar{\mathbf{y}} \in R^{n+k}} \sum_{i=1}^{n+k} \bar{\mathbf{H}}_i \bar{\mathbf{y}}_i$$
$$\text{subject to} \quad \bar{\mathbf{y}}_i \geq 0, 0 \leq i \leq n-k, n+1 \leq i \leq n+k$$
$$\sum_{i=n-k+1}^{n} \bar{\mathbf{y}}_i \geq \sum_{i=1}^{n-k} \bar{\mathbf{y}}_i$$
$$\sum_{i=1}^{n+k} \bar{\mathbf{y}}_i^2 \leq 1 \quad (104)$$

where $\bar{\mathbf{H}}_i$ is the $i$-th element of $\bar{\mathbf{H}}$. Let $\bar{\mathbf{z}} \in R^{n+k}$ be a vector such that $\bar{\mathbf{z}}_i = 1, 1 \leq i \leq n-k$, $\bar{\mathbf{z}}_i = -1, n-k+1 \leq i \leq n$, and $\bar{\mathbf{z}}_i = 0, n+1 \leq i \leq n+k$. One can then proceed in a fashion similar to the one from Subsection 4.1.1 and compute an upper bound based on duality. However, there will be two important differences. First, we now have $\bar{\mathbf{H}}$ instead of $\tilde{\mathbf{H}}$. Second we have $\bar{\mathbf{z}}$ instead of $\mathbf{z}$. One should, however, note that $\|\bar{\mathbf{z}}\|_2 = \|\mathbf{z}\|_2$. After repeating literally every step of the derivation from Subsection 4.1.1 one obtains the following analogue to equation (30)

$$w(\mathbf{h}, S_w) \leq \sqrt{\sum_{i=1}^{n+k} \bar{\mathbf{H}}_i^2 - \sum_{i=1}^{c} \bar{\mathbf{H}}_i^2 - \frac{((\bar{\mathbf{H}}^T \bar{\mathbf{z}}) - \sum_{i=1}^{c} \bar{\mathbf{H}}_i)^2}{n-c}} = \sqrt{\sum_{i=c+1}^{n+k} \bar{\mathbf{H}}_i^2 - \frac{((\bar{\mathbf{H}}^T \bar{\mathbf{z}}) - \sum_{i=1}^{c} \bar{\mathbf{H}}_i)^2}{n-c}}$$
(105)

where $c \leq (n-k)$ is such that $((\bar{\mathbf{H}}^T \bar{\mathbf{z}}) - \sum_{i=1}^{c} \bar{\mathbf{H}}_i) \geq 0$. As earlier, as long as $(\bar{\mathbf{H}}^T \bar{\mathbf{z}}) \geq 0$ there will be a $c$ (it is possible that $c = 0$) such that quantity on the most right hand side of (105) is an upper bound on $w(\mathbf{h}, S_w)$.

Using (105) we then establish the following analogue to Lemmas 1 and 4.

**Lemma 6.** *Let $\mathbf{h} \in R^{dn}$ be a vector with i.i.d. zero-mean unit variance gaussian components. Further let $\bar{\mathbf{H}}$ be as defined in (103) and $w(\mathbf{h}, S_w) = \max_{\mathbf{w} \in S_w}(\mathbf{h}^T \mathbf{w})$ where $S_w$ is as defined in (99). Let $\bar{\mathbf{z}} \in R^{n+k}$ be a vector such that $\bar{\mathbf{z}}_i = 1, 1 \leq i \leq n-k$, $\bar{\mathbf{z}}_i = -1, n-k+1 \leq i \leq n$, and $\bar{\mathbf{z}}_i = 0, n+1 \leq i \leq n+k$.*

*Then*

$$w(\mathbf{h}, S_w) \leq B_w \tag{106}$$

*where*

$$B_w = \begin{cases} \sqrt{\sum_{i=1}^{n+k} \bar{\mathbf{H}}_i^2} & \text{if } \zeta_w(\mathbf{h}, c_w) \leq 0 \\ \sqrt{\sum_{i=c_w+1}^{n+k} \bar{\mathbf{H}}_i^2 - \frac{((\bar{\mathbf{H}}^T\bar{\mathbf{z}}) - \sum_{i=1}^{c_w} \bar{\mathbf{H}}_i)^2}{n-c_w}} & \text{if } \zeta_w(\mathbf{h}, c_w) > 0 \end{cases}, \tag{107}$$

$\zeta_w(\mathbf{h}, c) = \frac{(\bar{\mathbf{H}}^T\bar{\mathbf{z}}) - \sum_{i=1}^{c} \bar{\mathbf{H}}_i}{n-c} - \bar{\mathbf{H}}_c$ and $c_w = \delta_w n$ is a $c \leq n-k$ such that

$$\frac{(1-\epsilon)E((\bar{\mathbf{H}}^T\bar{\mathbf{z}}) - \sum_{i=1}^{c} \bar{\mathbf{H}}_i)}{n-c} - F_{\chi_d}^{-1}\left(\frac{(1+\epsilon)c}{n-k}\right) = 0. \tag{108}$$

$F_{\chi_d}^{-1}(\cdot)$ *is the inverse cdf of the chi random variable with $d$ degrees of freedom. $\epsilon > 0$ is an arbitrarily small constant independent of $n$.*

*Proof.* Follows directly from the derivation before Lemma 1. $\square$

### 4.3.2 Computing an upper bound on $E(B_w)$

Following step-by-step the derivation of Lemma 3 (with a trivial adjustment in finding Lipschitz constant $\sigma$) we can establish the weak threshold analogue to it.

**Lemma 7.** *Assume the setup of Lemma 6. Let further $\psi_w = \frac{E(\bar{\mathbf{H}}^T\bar{\mathbf{z}}) - \sum_{i=1}^{c_w} \bar{\mathbf{H}}_i}{n}$. Then*

$$E(B_w) \leq \sqrt{n}\left(\exp\left\{-\frac{n\epsilon^2 \delta_w}{2(1+\epsilon)}\right\} + \exp\left\{-\frac{\epsilon^2 \psi_w^2 n}{2}\right\}\right) + \sqrt{E\sum_{i=c_w+1}^{n+k} \bar{\mathbf{H}}_i^2 - \frac{(E(\bar{\mathbf{H}}^T\bar{\mathbf{z}}) - E\sum_{i=1}^{c_w} \bar{\mathbf{H}}_i)^2}{n-c_w}}. \tag{109}$$

*Proof.* Follows directly from the derivation before Lemma 3. $\square$

Similarly to (50) and (83), if $n$ is large, for a fixed $\alpha$ one can determine $\beta_w$ as a maximum $\beta$ such that

$$\alpha d > \frac{E\sum_{i=c_w+1}^{n+k} \bar{\mathbf{H}}_i^2}{n} - \frac{(E(\bar{\mathbf{H}}^T\bar{\mathbf{z}}) - E\sum_{i=1}^{c_w} \bar{\mathbf{H}}_i)^2}{n(n-c_w)}. \tag{110}$$

In the rest of this subsection we show how the left hand side of (110) can be computed for a randomly chosen fixed $\beta_w$. We again, as earlier, do so in two steps:

1. We first determine $c_w$

2. We then compute $\lim_{n\to\infty}\left(\frac{E\sum_{i=c_w+1}^{n+k}\bar{\mathbf{H}}_i^2}{n} - \frac{(E(\bar{\mathbf{H}}^T\bar{\mathbf{z}})-E\sum_{i=1}^{c_w}\bar{\mathbf{H}}_i)^2}{n(n-c_w)}\right)$ with $c_w$ found in step 1.

*Step 1:*

From Lemma 6 we have $c_w = \delta_w n$ is a $c$ such that

$$\frac{(1-\epsilon)E((\sum_{i=1}^{n-\beta_w n}\bar{\mathbf{H}}_i - \sum_{i=n-\beta_w n+1}^{n+k}\bar{\mathbf{H}}_i) - \sum_{i=1}^{\delta_w n}\bar{\mathbf{H}}_i)}{n-c} - F_{\chi_d}^{-1}\left(\frac{(1+\epsilon)c}{n(1-\beta_w)}\right) = 0$$

$$\Leftrightarrow \frac{(1-\epsilon)(E\sum_{i=1}^{n-\beta_w n}\bar{\mathbf{H}}_i - E\sum_{i=n-\beta_w n+1}^{n}\bar{\mathbf{H}}_i - E\sum_{i=1}^{\delta_w n}\bar{\mathbf{H}}_i)}{n-c} - F_{\chi_d}^{-1}\left(\frac{(1+\epsilon)c}{n(1-\beta_w)}\right) = 0$$

$$\Leftrightarrow \frac{(1-\epsilon)(E\sum_{i=1}^{n-\beta_w n}\bar{\mathbf{H}}_i + E\sum_{i=n-\beta_w n+1}^{n}\mathbf{h}_{(i-1)d+1} - E\sum_{i=1}^{\delta_w n}\bar{\mathbf{H}}_i)}{n-c} - F_{\chi_d}^{-1}\left(\frac{(1+\epsilon)c}{n(1-\beta_w)}\right) = 0$$

$$\Leftrightarrow \frac{(1-\epsilon)(E\sum_{i=1}^{n-\beta_w n}\bar{\mathbf{H}}_i - E\sum_{i=1}^{\delta_w n}\bar{\mathbf{H}}_i)}{n-c} - F_{\chi_d}^{-1}\left(\frac{(1+\epsilon)c}{n(1-\beta_w)}\right) = 0$$

$$\Leftrightarrow \frac{(1-\epsilon)E\sum_{i=\delta_w n+1}^{n-\beta_w n}\bar{\mathbf{H}}_i}{n-c} - F_{\chi_d}^{-1}\left(\frac{(1+\epsilon)c}{n(1-\beta_w)}\right) = 0 \tag{111}$$

Set $\theta_w = 1 - \delta_w$. Then combining (111) and (86) we obtain the following equation for computing $\theta_w$

$$(1-\epsilon)(1-\beta_w)\frac{\frac{\sqrt{2}\Gamma(\frac{d+1}{2})}{\Gamma(\frac{d}{2})}\left(1 - \gamma_{inc}(\gamma_{inc}^{-1}(\frac{1-\theta_w}{1-\beta_w},\frac{d}{2}),\frac{d+1}{2})\right)}{\theta_w} - \sqrt{2\gamma_{inc}^{-1}(\frac{(1+\epsilon)(1-\theta_w)}{1-\beta_w},\frac{d}{2})} = 0. \tag{112}$$

Let $\hat{\theta}_w$ be the solution of (112). Then $\delta_w = 1 - \hat{\theta}_w$ and $c_w = \delta_w n = (1-\hat{\theta}_w)n$. This concludes step 1.

*Step 2:*

In this step we compute $\lim_{n\to\infty}\left(\frac{E\sum_{i=c_w+1}^{n+k}\bar{\mathbf{H}}_i^2}{n} - \frac{(E(\bar{\mathbf{H}}^T\bar{\mathbf{z}})-E\sum_{i=1}^{c_w}\bar{\mathbf{H}}_i)^2}{n(n-c_w)}\right)$ with $c_w = (1-\hat{\theta}_w)n$. Using results from step 1 we easily find

$$\lim_{n\to\infty}\frac{(E(\bar{\mathbf{H}}^T\bar{\mathbf{z}})-E\sum_{i=1}^{c_w}\bar{\mathbf{H}}_i)^2}{n(n-c_w)} = \frac{\left((1-\beta_w)\frac{\sqrt{2}\Gamma(\frac{d+1}{2})}{\Gamma(\frac{d}{2})}(1-\gamma_{inc}(\gamma_{inc}^{-1}(\frac{1-\hat{\theta}_w}{1-\beta_w},\frac{d}{2}),\frac{d+1}{2}))\right)^2}{\hat{\theta}_w}. \tag{113}$$

Effectively, what is left to compute is $\frac{E\sum_{i=c_w+1}^{n+k}\bar{\mathbf{H}}_i^2}{n}$. We first observe

$$
\begin{aligned}
\frac{E\sum_{i=c_w+1}^{n+k}\bar{\mathbf{H}}_i^2}{n} &= \frac{E\sum_{i=c_w+1}^{(1-\beta_w)n}\bar{\mathbf{H}}_i^2}{n} + \frac{E\sum_{i=(1-\beta_w)n+1}^{n}\bar{\mathbf{H}}_i^2}{n} + \frac{E\sum_{i=n+1}^{n+\beta_w n}\bar{\mathbf{H}}_i^2}{n} \\
&= \frac{E\sum_{i=(1-\hat{\theta}_w)n+1}^{(1-\beta_w)n}\bar{\mathbf{H}}_i^2}{n} + \frac{E\sum_{i=(1-\beta_w)n+1}^{n}\mathbf{h}_{(i-1)d+1}^2}{n} + \frac{E\sum_{i=n+1}^{n+\beta_w n}\|\mathbf{H}_i^*\|_2^2}{n} \\
&= \frac{E\sum_{i=(1-\hat{\theta}_w)n+1}^{(1-\beta_w)n}\bar{\mathbf{H}}_i^2}{n} + \frac{\beta_w n}{n} + \frac{\beta_w n(d-1)}{n} \\
&= \frac{E\sum_{i=(1-\hat{\theta}_w)n+1}^{(1-\beta_w)n}\bar{\mathbf{H}}_i^2}{n} + \beta_w d.
\end{aligned}
\quad (114)
$$

Combining (114) and (91) we find

$$
\lim_{n\to\infty}\frac{E\sum_{i=c_w+1}^{n+k}\bar{\mathbf{H}}_i^2}{n} = (1-\beta_w)\frac{2\Gamma(\frac{d+2}{2})}{\Gamma(\frac{d}{2})}\left(1-\gamma_{inc}(\gamma_{inc}^{-1}(\frac{1-\hat{\theta}_w}{1-\beta_w},\frac{d}{2}),\frac{d+2}{2})\right)+\beta_w d. \quad (115)
$$

We summarize the results from this section in the following theorem.

**Theorem 5.** *(Weak threshold) Let $A$ be a $dm \times dn$ measurement matrix in (1) with the null-space uniformly distributed in the Grassmanian. Let the unknown $\mathbf{x}$ in (1) be $k$-block-sparse with the length of its blocks $d$. Further, let the location and the directions of nonzero blocks of $\mathbf{x}$ be arbitrarily chosen but fixed. Let $k, m, n$ be large and let $\alpha = \frac{m}{n}$ and $\beta_w = \frac{k}{n}$ be constants independent of $m$ and $n$. Let $\gamma_{inc}(\cdot,\cdot)$ and $\gamma_{inc}^{-1}(\cdot,\cdot)$ be the incomplete gamma function and its inverse, respectively. Further, let $\epsilon > 0$ be an arbitrarily small constant and $\hat{\theta}_w$, $(\beta_w \le \hat{\theta}_w \le 1)$ be the solution of*

$$
(1-\epsilon)(1-\beta_w)\frac{\frac{\sqrt{2}\Gamma(\frac{d+1}{2})}{\Gamma(\frac{d}{2})}\left(1-\gamma_{inc}(\gamma_{inc}^{-1}(\frac{1-\hat{\theta}_w}{1-\beta_w},\frac{d}{2}),\frac{d+1}{2})\right)}{\hat{\theta}_w} - \sqrt{2\gamma_{inc}^{-1}(\frac{(1+\epsilon)(1-\hat{\theta}_w)}{1-\beta_w},\frac{d}{2})} = 0. \quad (116)
$$

*If $\alpha$ and $\beta_w$ further satisfy*

$$
\begin{aligned}
\alpha d > &(1-\beta_w)\frac{2\Gamma(\frac{d+2}{2})}{\Gamma(\frac{d}{2})}\left(1-\gamma_{inc}(\gamma_{inc}^{-1}(\frac{1-\hat{\theta}_w}{1-\beta_w},\frac{d}{2}),\frac{d+2}{2})\right)+\beta_w d \\
&- \frac{\left((1-\beta_w)\frac{\sqrt{2}\Gamma(\frac{d+1}{2})}{\Gamma(\frac{d}{2})}(1-\gamma_{inc}(\gamma_{inc}^{-1}(\frac{1-\hat{\theta}_w}{1-\beta_w},\frac{d}{2}),\frac{d+1}{2}))\right)^2}{\hat{\theta}_w}
\end{aligned}
\quad (117)
$$

*then the solutions of (1) and (3) coincide with overwhelming probability.*

*Proof.* Follows from the previous discussion combining (6), (101), (106), (109), (110), (112), (113), and

(115). □

The results for the weak threshold obtained from the above theorem for different block-lengths $d$ are presented on Figure 5. We also show on Figure 5 the results for $d \to \infty$ that we will discuss in more detail in the following section.

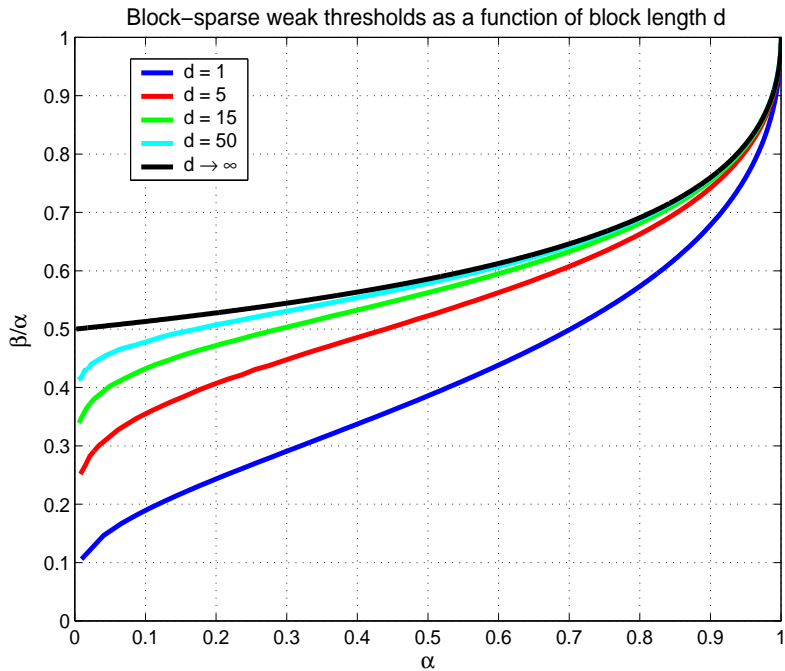

Figure 5: Block-sparse *weak* thresholds as a function of block length $d$, $\ell_2/\ell_1$-optimization

## 5   $d \to \infty$

When the block length is large one can simplify the conditions for finding the thresholds obtained in the previous section. Hence, in this section we establish attainable strong, sectional, and weak thresholds when $d \to \infty$, i.e. we establish attainable ultimate benefit of $\ell_2/\ell_1$-optimization from (3) when used in block-sparse recovery (1). Throughout this section we choose $d \to \infty$ in order to simplify the exposition. However, as it will become obvious, the analogous simplified expressions can in fact be obtained for any value of $d$.

### 5.1   $d \to \infty$ – strong threshold

Following the derivation of Section 4.1.1 and its connection to Theorem 3 it is not that difficult to see that choosing $\hat{\theta}_s = 1$ in (69) would provide a valid threshold condition as well ($\hat{\theta}_s = 1$ is in general not optimal for a fixed value $d$, i.e. when $d$ is not large a better choice for $\hat{\theta}_s$ is the one given in Theorem 3). The choice $\hat{\theta}_s = 1$ gives us the following corollary of Theorem 3.

**Corollary 3.** *(Strong threshold, $d \to \infty$) Let $A$ be a $dm \times dn$ measurement matrix in (1) with the null-space uniformly distributed in the Grassmanian. Let the unknown $\mathbf{x}$ in (1) be $k$-block-sparse with the length of its blocks $d \to \infty$. Let $k, m, n$ be large and let $\alpha = \frac{m}{n}$ and $\beta_s^\infty = \frac{k}{n}$ be constants independent of $m$ and $n$. Assume that $d$ is independent of $n$. If $\alpha$ and $\beta_s^\infty$ satisfy*

$$\alpha > 4\beta_s^\infty (1 - \beta_s^\infty) \tag{118}$$

*then the solutions of (1) and (3) coincide with overwhelming probability.*

*Proof.* Let $\hat{\theta}_s = 1$ in (69). Then from (69) we have

$$\begin{aligned}
\alpha &> \frac{2\Gamma(\frac{d+2}{2})}{d\Gamma(\frac{d}{2})} - \frac{\left(\frac{\sqrt{2}\Gamma(\frac{d+1}{2})}{\Gamma(\frac{d}{2})}(1 - 2(1 - \gamma_{inc}(\gamma_{inc}^{-1}(1 - \beta_s, \frac{d}{2}), \frac{d+1}{2})))\right)^2}{d} \\
&= 1 - \left((1 - 2(1 - \gamma_{inc}(\gamma_{inc}^{-1}(1 - \beta_s, \frac{d}{2}), \frac{d+1}{2})))\right)^2 \frac{\left(\frac{\sqrt{2}\Gamma(\frac{d+1}{2})}{\Gamma(\frac{d}{2})}\right)^2}{d}.
\end{aligned} \tag{119}$$

When $d \to \infty$ we have $\lim_{d \to \infty} \gamma_{inc}(\gamma_{inc}^{-1}(1 - \beta_s, \frac{d}{2}), \frac{d+1}{2})) = 1 - \beta_s$ and $\lim_{d \to \infty} \frac{1}{d}\left(\frac{\sqrt{2}\Gamma(\frac{d+1}{2})}{\Gamma\frac{d}{2}}\right)^2 = 1$. Then from (119) we obtain the following condition

$$\alpha > 1 - (1 - 2(1 - (1 - \beta_s)))^2 = 4\beta_s(1 - \beta_s). \tag{120}$$

Since (120) is exactly the same as (118) this concludes the proof. $\square$

The results obtained in the previous corollary precisely match those obtained in [78, 81].

## 5.2  $d \to \infty$ – sectional threshold

Following the derivation of Section 4.1.1 and its connection to Theorem 4 it is not that difficult to see that choosing $\hat{\theta}_{sec} = 1$ in (94) would provide a valid threshold condition as well (again, $\hat{\theta}_{sec} = 1$ is in general not optimal for a fixed value $d$, i.e. when $d$ is not large a better choice for $\hat{\theta}_{sec}$ is the one given in Theorem 4). Choosing $\hat{\theta}_{sec} = 1$ in (94) gives us the following corollary of Theorem 4.

**Corollary 4.** *(Sectional threshold, $d \to \infty$) Let $A$ be a $dm \times dn$ measurement matrix in (1) with the null-space uniformly distributed in the Grassmanian. Let the unknown $\mathbf{x}$ in (1) be $k$-block-sparse with the length of its blocks $d \to \infty$. Further, let the location of nonzero blocks of $\mathbf{x}$ be arbitrarily chosen but fixed. Let $k, m, n$ be large and let $\alpha = \frac{m}{n}$ and $\beta_{sec}^{\infty} = \frac{k}{n}$ be constants independent of $m$ and $n$. Assume that $d$ is independent of $n$. If $\alpha$ and $\beta_{sec}^{\infty}$ satisfy*

$$\alpha > 4\beta_{sec}^{\infty}(1 - \beta_{sec}^{\infty}) \tag{121}$$

*then the solutions of (1) and (3) coincide with overwhelming probability.*

*Proof.* Let $\hat{\theta}_{sec} = 1$ in (94). Then from (94) we have

$$\begin{aligned}
\alpha &> \frac{(1-\beta_{sec})d + \beta_{sec}d}{d} - \frac{\left((1-\beta_{sec})\frac{\sqrt{2}\Gamma(\frac{d+1}{2})}{\Gamma(\frac{d}{2})} - \frac{\sqrt{2}\Gamma(\frac{d+1}{2})}{\Gamma(\frac{d}{2})}\beta_{sec}\right)^2}{d} \\
&= 1 - (1 - 2\beta_{sec})^2 \frac{1}{d}\left(\frac{\sqrt{2}\Gamma(\frac{d+1}{2})}{\Gamma(\frac{d}{2})}\right)^2.
\end{aligned} \tag{122}$$

When $d \to \infty$ we have $\lim_{d \to \infty} \frac{1}{d}\left(\frac{\sqrt{2}\Gamma(\frac{d+1}{2})}{\Gamma(\frac{d}{2})}\right)^2 = 1$. Then from (122) we easily obtain the condition

$$\alpha > 4\beta_{sec}(1 - \beta_{sec})$$

which is the same as the condition stated in (121). This therefore concludes the proof. □

**Remark:** Of course, the statement of Corollary 4 could have been deduced trivially from Corollary 3. Namely, any attainable value of the strong threshold is an attainable value for the sectional threshold as well.

### 5.3  $d \to \infty$ – weak threshold

Reasoning as in the two previous subsections we have that $\hat{\theta}_w = 1$ in (117) would provide a valid condition for computing the weak threshold. In turn choosing $\hat{\theta}_w = 1$ in (117) gives us the following corollary of Theorem 5.

**Corollary 5.** *(Weak threshold, $d \to \infty$) Let $A$ be a $dm \times dn$ measurement matrix in (1) with the null-space uniformly distributed in the Grassmanian. Let the unknown $\mathbf{x}$ in (1) be $k$-block-sparse with the length of its blocks $d \to \infty$ Further, let the location and the directions of nonzero blocks of $\mathbf{x}$ be arbitrarily chosen but fixed. Let $k, m, n$ be large and let $\alpha = \frac{m}{n}$ and $\beta_w^{\infty} = \frac{k}{n}$ be constants independent of $m$ and $n$. Assume that*

*$d$ is independent of $n$. If $\alpha$ and $\beta_w^\infty$ satisfy*

$$\alpha > \beta_w^\infty(2 - \beta_w^\infty) \tag{123}$$

*then the solutions of (1) and (3) coincide with overwhelming probability.*

*Proof.* Let $\hat{\theta}_w = 1$ in (117). Then from (117) we have

$$\begin{aligned}
\alpha &> \frac{(1-\beta_w)d + \beta_w d}{d} - \frac{\left((1-\beta_w)\frac{\sqrt{2}\Gamma(\frac{d+1}{2})}{\Gamma(\frac{d}{2})}\right)^2}{d} \\
&= 1 - \frac{\left((1-\beta_w)\frac{\sqrt{2}\Gamma(\frac{d+1}{2})}{\Gamma(\frac{d}{2})}\right)^2}{d}.
\end{aligned} \tag{124}$$

As earlier, when $d \to \infty$ we have $\lim_{d \to \infty} \frac{1}{d}\left(\frac{\sqrt{2}\Gamma(\frac{d+1}{2})}{\Gamma(\frac{d}{2})}\right)^2 = 1$. Then from (124) we easily obtain the condition

$$\alpha > \beta_w(2 - \beta_w)$$

which is the same as the condition stated in (123). This therefore concludes the proof. □

The results for the strong, sectional, and weak threshold obtained in the three above corollaries are shown on figures in earlier sections as curves denoted by $d \to \infty$.

It is interesting to note that (119), (122), and (124) can be used instead of (69), (94), and (117) to determine attainable values of the thresholds for any fixed $d$. Given that (119), (122), and (124) are obtained for a suboptimal choice of $\hat{\theta}$ the threshold values that they produce trail those presented on Figures 3, 4, and 5 and we therefore do not include them in this paper. However, we do mention that they are relatively easier to compute and a fairly good approximation of the results presented on Figures 3, 4, and 5.

## 6 Numerical experiments

In this section we briefly discuss the results that we obtained from numerical experiments. In all our numerical experiments we fixed $n = 100$ and $d = 15$. We then generated matrices $A$ of size $dm \times dn$ with $m = (10, 20, 30, \ldots, 90, 99)$. The components of the measurement matrices $A$ were generated as i.i.d. zero-mean unit variance Gaussian random variables. For each $m$ we generated $k$-block-sparse signals $\mathbf{x}$ for several different values of $k$ from the transition zone (the locations of non-zero blocks of $\mathbf{x}$ were chosen randomly). For each combination $(k, m)$ we generated 100 different problem instances and recorded the number of times $\ell_2/\ell_1$-optimization algorithm from (3) failed to recover the correct $k$-block-sparse $\mathbf{x}$. All different $(k, m)$ combinations as well as the corresponding numbers of failed experiments are given in Table 1. The interpolated data from Table 1 are presented graphically on Figure 6. The color of any point on Figure 6 shows the probability of having $\ell_2/\ell_1$-optimization succeed for a combination $(\alpha, \beta)$ that corresponds to that point. The colors are mapped to probabilities according to the scale on the right hand side of the figure. The simulated results can naturally be compared to the weak threshold theoretical prediction. Hence, we also show on Figure 6 the theoretical value for the weak threshold calculated according to Theorem 5 (and shown on Figure 5). We observe that the simulation results are in a good agreement with the theoretical calculation.

Table 1: The simulation results for recovery of block-sparse signals; $n = 100$, $d = 15$

| $m$ | 10 | 20 | 30 | 40 | 50 | 60 | 70 | 80 | 90 | 99 |
|---|---|---|---|---|---|---|---|---|---|---|
| $k$ / # of errors | 7/100 | 12/100 | 18/100 | 22/76 | 29/80 | 37/94 | 46/95 | 57/98 | 71/97 | 92/89 |
| $k$ / # of errors | 6/100 | 11/98 | 17/100 | 22/76 | 29/80 | 36/64 | 45/71 | 55/60 | 69/70 | 90/52 |
| $k$ / # of errors | 5/95 | 10/93 | 16/89 | 21/39 | 28/43 | 35/26 | 44/38 | 53/11 | 67/27 | 89/27 |
| $k$ / # of errors | 4/14 | 9/21 | 15/36 | 20/5 | 27/11 | 34/6 | 43/11 | 52/2 | 66/11 | 88/12 |
| $k$ / # of errors | 3/0 | 8/0 | 14/8 | 19/0 | 25/0 | 32/0 | 42/6 | 50/0 | 65/6 | 87/3 |

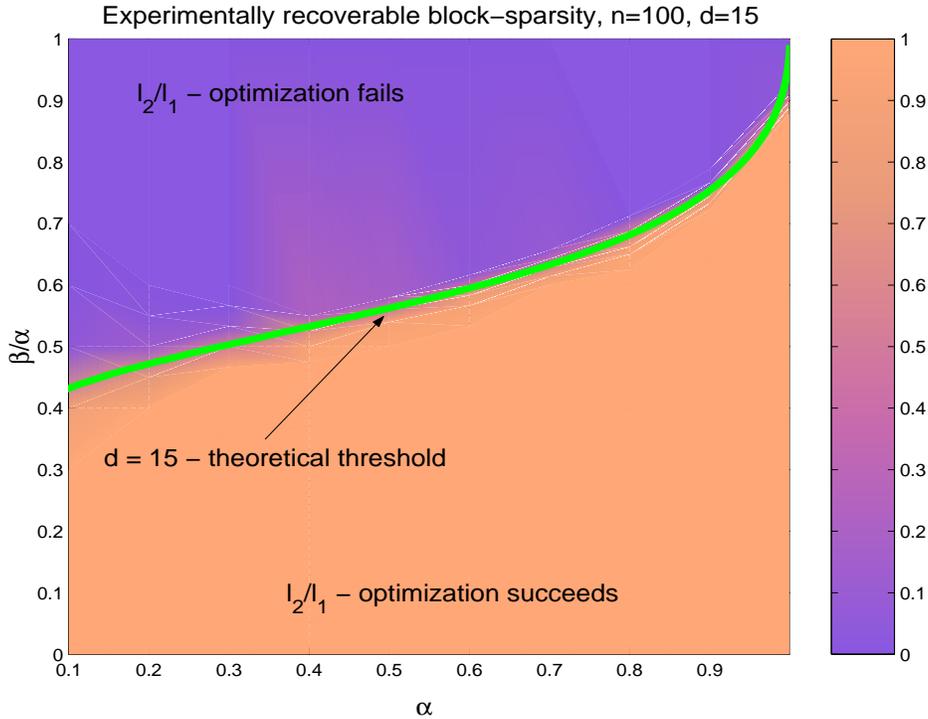

Figure 6: Experimentally recoverable block-sparsity, $\ell_2/\ell_1$-optimization

# 7  Discussion

In this paper we considered recovery of block-sparse signals from a reduced number of linear measurements. We provided a theoretical performance analysis of a polynomial $\ell_2/\ell_1$-optimization algorithm. Under the assumption that the measurement matrix $A$ has a basis of the null-space distributed uniformly in the Grassmanian, we derived lower bounds on the values of the recoverable strong, sectional, and weak thresholds in the so-called linear regime, i.e. in the regime when the recoverable sparsity is proportional to the length of the unknown vector. We also conducted the numerical experiments and observed a solid agreement between the simulated and the theoretical weak threshold.

The main subject of this paper was the recovery of the so-called ideally block-sparse signals. However, the presented analysis framework admits various generalizations. Namely, it can be extended to include computations of threshold values for recovery of approximately block-sparse signals as well as those with noisy measurements. Also, in this paper we were mostly concerned with the success of $\ell_2/\ell_1$-optimization. However, as we have mentioned earlier instead of $\ell_2/\ell_1$-optimization one could use an $\ell_2/\ell_q$-optimization ($0<q<1$). While the resulting problem would not be convex it could still be solved (not necessarily in polynomial time) with various techniques from the literature. One could then potentially find an interest in generalizing the results of the present paper to the case of $\ell_2/\ell_q$-optimization ($0<q<1$) as well. On a completely different note, carefully following our exposition one could spot that the results presented in this paper assume large dimensions of the system. Obtaining their equivalents for systems of moderate dimensions is another possible generalization. All these generalizations will be part of a future work.

We would like to reemphasize that our analysis heavily relied on a particular probability distribution of the null-space of the measurement matrix. On the other hand our extensive numerical experiments (results of some of them are presented in [83]) indicate that $\ell_2/\ell_1$-optimization works equally well for many different statistical measurement matrices $A$ (e.g. Bernoulli). It will be interesting to see if the analysis presented here can be generalized to these cases as well. Furthermore, as in [33], one can raise the question of identifying class of statistical matrices for which $\ell_2/\ell_1$-optimization works as well as in the case presented in this paper. However, we do believe that answering this question is not an easy task.

As far as the technical contribution goes, we should mention that our analysis made a critical use of an excellent work [47] which on the other hand massively relied on phenomenal results [20, 67] related to the estimates of the normal tail distributions of Lipshitz functions. In a very recent work related to the matrix-rank optimization the authors in [69] successfully conducted a theoretical analysis applying results of [20,67] without relying on the conclusions of [47]. It will certainly be interesting to see what performance guarantees the direct application of the results of [20, 67] would produce in the problems considered in this paper.

Lastly, it is relatively easy to note that the signal structure imposed in this paper is very simple, i.e. almost ideal. For example, we assumed that all blocks are of the same length. Just slightly modifying that assumption so that the blocks are not of equal length significantly complicates the problem. It will be interesting to see if algorithms similar to $\ell_2/\ell_1$-optimization can be used for signals with these (or possibly even some completely different) structures and if an analysis similar to the one presented in this paper can be developed for them as well.